# Ion implantation in β-$Ga_2O_3$: physics and technology

Alena Nikolskaya [1)], Evgenia Okulich [1)], Dmitry Korolev [1)], Anton Stepanov [2)], Dmitry Nikolichev [1)], Alexey Mikhaylov [1)], David Tetelbaum [1)], Aleksei Almaev [3)], Charles Airton Bolzan [4)], Antônio Jr Buaczik [4)], Raquel Giulian [4)], Pedro Luis Grande [4)], Ashok Kumar [5)], Mahesh Kumar [5)], Daniela Gogova [1,6,a)]

[1]Research Institute of Physics and Technology, Lobachevsky University, Nizhny Novgorod 603950, Russia
[2]Chuvash State Agricultural Academy, Cheboksary, 428017 Russia
[3]Research and Development Center for Advanced Technologies in Microelectronics, Tomsk State University, Tomsk, 634050, Russia
[4]Ion Implantation Laboratory, Institute of Physics - Federal University of Rio Grande do Sul, Brazil
[5]Department of Electrical Engineering, Indian Institute of Technology Jodhpur, Jodhpur, India
[6]Centre for Materials Science and Nanotechnology, University of Oslo, Blindern, 0316, Oslo, Norway

[a)] Electronic mail: daniela.gogova-petrova@smn.uio.no

Gallium oxide and in particular its thermodynamically stable β-$Ga_2O_3$ phase is within the most exciting materials in research and technology nowadays due to its unique properties, such as an ultra-wide band gap and a very high breakdown electric field, finding a number of applications in electronics and optoelectronics. Ion implantation is a traditional technological method used in these fields, and its well-known advantages can contribute greatly to the rapid development of physics and technology of $Ga_2O_3$-based materials and devices. Here, the current status of ion beam implantation in β-$Ga_2O_3$ is reviewed. The main attention is paid to the results of experimental study of damage under ion irradiation and the properties of $Ga_2O_3$ layers doped by ion implantation. The results of *ab initio* theoretical calculations of the impurities and defects parameters are briefly presented, and the physical principles of a number of analytical methods used to study implanted gallium oxide layers are highlighted. The use of ion implantation in the development of such $Ga_2O_3$-based devices as metal oxide field effect transistors, Schottky barrier diodes, and solar-blind UV detectors, is described together with systematical analysis of the achieved values of their characteristics. Finally, the most important challenges to be overcome in this field of science and technology are discussed.

# I. INTRODUCTION

Gallium oxide ($Ga_2O_3$) is an ultra-wide band gap semiconductor that has attracted enormous attention from researchers and engineers in the field of solid-state electronics. The history of research on this material goes back more than over 60 years, but only recently it has acquired the reputation of a candidate for one of the main material of advanced electronic and optoelectronic devices, such as power transistors and high-voltage diodes, solar blind UV photodetectors, gas sensors, etc. This is due to the growing demands for such devices.

Initially, the attention of researchers was focused on such wide band gap materials as SiC and GaN. However, the situation changed after high-quality power field-effect transistors were created on the basis of β-$Ga_2O_3$, which is the most stable modification of gallium oxide. Since then, the number of publications devoted to $Ga_2O_3$ has a steep surge (Figure 1), including several review articles [1,2,3,4,5,6,7].

The main advantages of this material are the following:

1) The wide bandgap (4.5 – 4.9 eV) and the very high breakdown electric field (about 8 MeV/cm, according to theoretical calculations), which is larger than the corresponding values of SiC and GaN.

2) Radiation hardness and chemical resistance, allowing the devices to remain operational under conditions of ionizing radiation and in aggressive environment.

3) Existence of the melt growth techniques of large-diameter ingots in comparison to the expensive methods of SiC and GaN crystals fabrication.

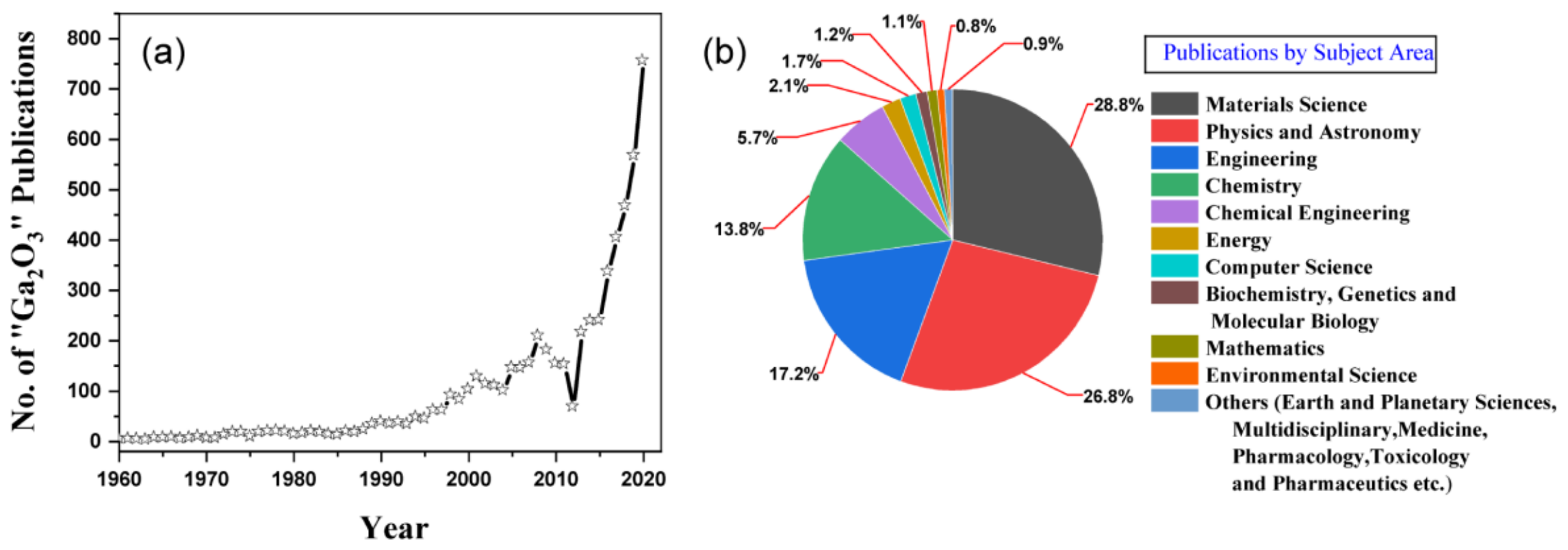

Figure 1. Number of publications on $Ga_2O_3$ from 1960 to 2020 (a) and their distribution over subject areas (b). The papers were searched with the criterion of containing “$Ga_2O_3$” in the title (Data: Scopus)

It is known that electronics has reached great success thanks to the discovery of ion implantation as a key method in processing of practically all semiconductor devices. The main advantages of this method include: 1) the precise dosage of dopants; 2) ability to control the spatial distribution of dopants and to use radiation defects generated by ion irradiation to modify the material properties - "defect engineering"; 3) relatively low temperatures required for the implantation-related technology. All these advantages and a number of other useful features of ion implantation can be exploited to enhance the gallium oxide potential in device applications. The significance of ion implantation additionally increases due to the well-known problem of obtaining $Ga_2O_3$ *p*-type conductivity under growth conditions close to thermodynamic equilibrium. The ion implantation as a non-equilibrium method seems to be a promising approach to overcome this challenge.

Investigations on the $Ga_2O_3$ ion implantation have begun quite recently, and only a relatively small circle of fundamental or applied problems has been faced so far. Nevertheless, the advantages of this method have already been demonstrated for a number of important practical applications. In particular, the possibility of a significant improvement in the properties of ohmic contacts by ion doping is established, which makes it possible to enhance the operating parameters of some semiconductor devices. It is also important that the possibility of a controlled increase in the resistivity of *n*-$Ga_2O_3$ by ion doping by some acceptor impurities and radiation defects incorporation is discovered. New approaches are developed for the ion synthesis and ion modification of $Ga_2O_3$ nanostructures for optoelectronic devices. Meanwhile, successful developments in material science and engineering are impossible without thorough theoretical modeling. For research in the area of ion implantation, it is necessary to use the results of computer calculations regarding the influence of impurities and point defects on the electron structure and properties of β-$Ga_2O_3$.

This article is devoted to a comprehensive review of the main published results in the field of β-$Ga_2O_3$ ion implantation.

The review is organized as follows. After the Introduction (Section I), the key results on computer simulations of the electronic structure of β-$Ga_2O_3$ containing dopants and defects are reviewed (Section II). Section III describes the main characterization methods employed in the experiments with ion implantation of $Ga_2O_3$. This section is required predominantly for readers who are not experts in materials characterization. In the next Section IV, the properties of impurities in β-$Ga_2O_3$ introduced by ion implantation are examined. Then, the features of defect formation upon $Ga_2O_3$ irradiation with ions of different masses and energies are briefly reviewed. In Section V, both scientific and applied aspects related to the development of $Ga_2O_3$-based electronic and optoelectronic devices are highlighted. Finally, in the Conclusions section, a general assessment of the current research state-of-the-art and the main challenges to be addressed in the field of β-$Ga_2O_3$ ion implantation are given.

## II. THEORETICAL CALCULATIONS

The study of physical foundations of ion implantation is generally aimed to solve the following three problems: (a) Radiation defects: their nature, regularities of formation and annealing, distributions, diffusion rates, influence on the properties of implanted layers; (b) Ion doping regularities: the implanted impurities activation processes, energy levels, diffusion rates, distributions, influence on electrical, optical and other properties; (c) Defect-impurity interaction: the influence of interaction on the diffusion rates and on the implanted layer properties. The term "defects", where this doesn't lead to confusion, will be used below to denote both exactly defects and impurities, as well.

Although the modern arsenal of experimental methods provides rich opportunities for solving the mentioned problems, some of them often cannot be reliably solved without theoretical consideration, especially without the *ab initio* methods. Among such problems are the identification of defects and impurities responsible for one or another experimentally found energy levels, explanation of the observed properties.

At present, almost all *ab initio* calculations are limited to consideration of the elementary entities – vacancies, interstitials, single impurity atoms, and some simple complexes. Of course, for ion implantation, the list of formed defects types is significantly greater and includes complicated complexes, extended defects, disordered nanoregions, etc. However, in some cases, e.g., upon irradiation with light ions at not too high doses, mainly the same types of defects are present as those introduced during crystal growth or epitaxy. Thus, the knowledge of theoretical approaches and the results of corresponding calculations is very desirable for researchers dealing with ion implantation.

The basic *ab initio* approach most often used to study β-$Ga_2O_3$ materials is the so-called Density-Function Theory (DFT). The basics of the approach are considered, for example, in Ref. [8] [9]. For a wide band gap semiconductor (with strongly expressed ionic type of atomic bonding), there are some peculiarities that need to be taken into account when using this approach. Thus, the standard Local Density Approximation (LDA) and the Semilocal Generalized Gradient Approximation (GGA) are not well adequate [10]. In modern works, to take into account these peculiarities, the so-called hybrid exchange-density functional is used [11] [12]. For β-$Ga_2O_3$, this is the HSE06 functional [13] [14] [15] [16] [17]. The choice of optimal parameters included in the HSE06 functional to give the best agreement with the experimental value of the β-$Ga_2O_3$ band gap was proposed in Ref. [10], although some other parameters are used, too. The alternative *ab initio* approach is the (GGA+U) with exchange-correlation energy PBE [18].

The DFT-calculated values of levels formed for typical defects and impurities in β-$Ga_2O_3$, as well as respective experimental data are presented in Table 1. The data of different theoretical works do not always agree with each other due to the difference in the parameters used, and in some cases because of the different calculation methods. However, the qualitative conclusions drawn by different authors are generally identical.

The comparison of the calculated and experimental values of the levels is only possible if they are correctly attributed to one or other entities. Authors of experimental works usually justify their attribution by the proximity of the found values of levels to one or another theoretically calculated value. However, optical or electrical methods do not allow direct attribution by themselves. In cases where defects and impurities are introduced during growth, the attribution problem is facilitated by the fact that the resulting state is often (although not always) close to thermodynamic equilibrium, so it is possible to take into account the calculated quantity of the defect formation energies: the smaller the formation energy, the more energetically favorable the formation of a given defect is. However, in the case of ion implantation, the state of implanted layers is generally far from equilibrium. Although the system approaches equilibrium upon annealing, the degree of approach strongly depends on the annealing conditions, and is not always known. Thus, for ion implantation, the attribution of defects is a more difficult task and requires some additional assumptions, including the estimation of kinetic factors (see e.g. Ref. [19]).

Table 1. The values of DFT-calculated energy levels of defects, impurities and STHs

| Defect, impurity, STH | Charge transition | Energy level, eV | Kind of DFT approach, dielectric constant | Reference | Experimental value |
|---|---|---|---|---|---|
| $V_{GaI}$ | -2/-3 | $E_C$ – 0.67 | HSE, $\varepsilon_\infty$ | Ref. [10] | $E_C$ – 0.74 Ref. [20]<br>$E_C$ – 0.82 Ref. [21] |
| | | $E_C$ – 1.62 | HSE, $\varepsilon_0$ | Ref. [15] | |
| | | $E_C$ – 1.76 | HSE, $\varepsilon_0$ | Ref. [19] | |
| | | $E_C$ – 0.69 | HSE, $\varepsilon_\infty$ | | |
| | | $E_C$ – 2.27 | $\varepsilon$ = 12.7 | Ref. [22] | |
| $V_{GaII}$ | -2/-3 | $E_C$ – 1.16 | HSE, $\varepsilon_\infty$ | Ref. [10] | $E_C$ – 1.04 Ref. [20]<br>$E_C$ – 1.00 Ref. [21] |
| | | $E_C$ – 1.83 | HSE, $\varepsilon_0$ | Ref. [15] | |
| | | $E_C$ – 1.93 | $\varepsilon$ = 12.7 | Ref. [22] | |
| | | $E_C$ – 2.17 | HSE, $\varepsilon_0$ | Ref. [19] | |
| | | $E_C$ – 1.11 | HSE, $\varepsilon_\infty$ | | |
| $V_{OI}$ | +2/0 | $E_C$ – 2.10 | HSE, $\varepsilon_\infty$ | Ref. [10] | |
| | | $E_C$ – 1.52 | HSE, $\varepsilon_0$ | Ref. [17] | |
| | | $E_C - 1.72$ | PBE, (GGA+U) approach | Ref. [23] | |
| | | $E_C - 1.50$ | HSE, $\varepsilon_0$ | Ref. [19] | |
| $V_{OII}$ | +2/0 | $E_C - 2.68$ | HSE, $\varepsilon_\infty$ | Ref. [10] | |
| | | $E_C - 2.29$ | HSE, $\varepsilon_0$ | | $E_C - 2.16$ Ref. [21] |
| | | $E_C - 2.16$ | HSE, $\varepsilon_0$ | Ref. [17] | |
| | | $E_C - 2.42$ | $\varepsilon = 12.7$ | Ref. [22] | |
| | | $E_C - 1.22$ | PBE, (GGA+U) approach | Ref. [23] | $E_C - 1.40$ Ref. [23] |
| | | $E_C - 2.23$ | HSE, $\varepsilon_0$ | Ref. [19] | |
| $V_{OIII}$ | +2/0 | $E_C - 1.95$ | HSE, $\varepsilon_\infty$ | Ref. [10] | |
| | | $E_C - 1.26$ | HSE, $\varepsilon_0$ | Ref. [17] | |
| | | $E_C - 1.02$ | PBE, (GGA+U) approach | Ref. [23] | |
| | | $E_C - 1.36$ | HSE, $\varepsilon_0$ | Ref. [19] | |
| | | $E_C - 1.79$ | HSE, $\varepsilon_\infty$ | Ref. [19] | |
| $Mg_{GaI}$ | 0/-1 | $E_V + 1.25$ | | Ref. [24] | |
| | | $E_V + 1.27$ | HSE, $\varepsilon_0$ | Ref. [25] | |
| | | $E_V + 1.62$ | | Ref. [26] | |
| | | $E_V + 1.30$ | | Ref. [27] | |
| $Mg_{GaII}$ | 0/-1 | $E_V + 1.05$ | | Ref. [24] | $E_V + 1.1$ Ref. [28] |
| | | $E_V + 1.06$ | HSE, $\varepsilon_0$ | Ref. [25] | |
| | | $E_V + 1.25$ | HSE, $\varepsilon_0$ | Ref. [29] | |
| | | $E_V + 1.57$ | | Ref. [26] | |
| | | $E_V + 1.40$ | | Ref. [27] | |
| | | $E_V + 1.40$ | | Ref. [30] | |
| $N_{OI}$ | 0/-1 | $E_V + 3.50$ | HSE, $\varepsilon_0$ | Ref. [29] | |
| | | $E_V + 3.50$ | | Ref. [27] | |
| | | $E_V + 3.40$ | | Ref. [30] | |
| $N_{OII}$ | 0/-1 | $E_V + 3.40$ | HSE, $\varepsilon_0$ | Ref. [29] | |
| | | $E_V + 2.20$ | | Ref. [27] | |
| | | $E_V$ + 3.50 | | Ref. [30] | |
| $N_{OIII}$ | 0/-1 | $E_V$ + 2.20 | HSE, $\varepsilon_0$ | Ref. [29] | |
| | | $E_V$ + 3.40 | | Ref. [27] | |
| | | $E_V$ + 2.20 | | Ref. [30] | |
| $V_{GaI}$-H | +1/0 | $E_C$ – 0.89 | | Ref. [31] | |
| | 0/-1 | $E_C$ – 1.53 | | | |
| | -1/-2 | $E_C$ – 2.21 | | | |
| | | $E_C$ – 1.93 | | Ref. [15] | |
| $V_{GaI}$-2H | +1/0 | $E_C$ – 1.90 | | Ref. [31] | |
| | 0/-1 | $E_C$ – 1.64 | | | |
| | -1/-2 | $E_C$ – 1.77 | | | |
| $V_{GaII}$-3H | +1/0 | $E_C$ – 1.28 | | Ref. [31] | |
| | 0/-1 | $E_C$ – 1.15 | | | |
| | -1/-2 | $E_C$ – 2.25 | | | |
| $Fe_{GaI}$ | 0/-1 | $E_C$ – 0.59 | | Ref. [19] | |
| $STH_{OI}$ | +1/0 | $E_V$ + 0.58 | | Ref. [27] | |
| | | $E_C$ – 4.61 | | Ref. [10] | |
| $STH_{OII}$ | +1/0 | $E_C$ – 4.50 | | Ref. [10] | |
| $STH_{OIII}$ | +1/0 | $E_V$ + 0.58 | | Ref. [27] | |

The notations: $E_C$ – conduction band edge, $E_V$ – valence band edge, $\varepsilon_0 \approx 10$ – static dielectric constant, $\varepsilon_\infty \approx 3.55$ – high frequency dielectric constant, V – vacancy, $Ga_I$, $Ga_{II}$, $O_I$, $O_{II}$, $O_{III}$ correspond to different lattice positions of Ga and O atoms, $STH_{OI}$, $STH_{OII}$ and $STH_{OIII}$ correspond to the respective positions of trapped hole – at the O atoms.

In addition to defects and impurities, the DFT method was used to calculate the energy levels of the so-called "Self Trapped Holes (STH)". STH appears when the free hole state (in the valence band) is less energetically favorable than for its localization near some lattice positions [16]. For β-$Ga_2O_3$, according to calculations, these positions are localized near the oxygen atoms [16]. They have three nonequivalent positions in the unit / cell [3] and correspondingly in general case create three STH levels in the band gap (although some of them, as shown in Ref. [27], may be unstable). The localization of a hole at the STH level causes a strong distortion of the lattice in its vicinity that leads to the formation of the so-called small polaron [32] which, according to calculations [16], has an exceedingly low mobility. This circumstance is considered as one (but not single) of the main challenges on the way to *p*-type β-$Ga_2O_3$.

The existence of STH in β-$Ga_2O_3$ was confirmed in Ref. [33] by the photoluminescence method with polarized light excitation. The STH levels were calculated by DFT [10 16 27 33] and found in the experimental work [34].

First-principal calculations were also used to calculate the migration energies of defects and impurities in β-$Ga_2O_3$. Thus, in Ref. [35], the barriers for migration of $V_O^{2+}$ and $V_O^0$ were calculated to be – 1.8 and 2.6 eV, respectively. For $V_{Ga}$, lower barrier values were obtained and a conclusion was drawn about their higher mobility at low temperatures, compared to the $V_O$.

In Ref. [30], the energies of Mg migration over interstitial sites (from 0.56 to 0.75 eV) and N for oxygen vacancies were calculated as 3.87 eV.

The results presented show that theoretical study on defects and impurities in β-$Ga_2O_3$ requires further development and experimental verification. In particular, it is not entirely clear which value of ε ($\varepsilon_0$ or $\varepsilon_\infty$) gives more correct results when using the HSE functional. With increase in the size of computational cell, the results calculated with $\varepsilon_0$ apparently will become more reliable, but this increase will require a significant elevation in the computer resources. One more question is also not entirely resolved which of the two approaches (HSE or GGA+U) provides the best agreement with the experimental data. Nevertheless, already at the present stage, the use of theoretical calculations in the interpretation of experimental results is very useful and even it is often mandatory.

# III. ANALYTICAL METHODS

The analytical methods most often used in the study of $Ga_2O_3$ are discussed below.

*Transmission Electron Microscopy*

The method of transmission electron microscopy (TEM) serves to study the crystalline structure and extended defects of thin films or thinned layers of bulk samples[36]. This very powerful method is based on the use of dual – corpuscular and wave nature of electrons. The electrons of the parallel incident beam passing through the sample (with energies usually of the order of several hundred keV) undergo diffraction on the crystal lattice (coherent scattering) and/or incoherent scattering on atoms. Transmitted and scattered electrons are controlled using electronic lens systems and variable aperture diaphragms (Figure 2). So, the images and diffraction pattern for the selected area of a sample can be formed.

The diffraction mode allows determining the nature of the crystalline structure of the sample studied and identifying the material by comparison with reference data. A single-crystalline material forms a system of point reflections, a polycrystalline one – a system of narrow rings and an amorphous film forms a halo or a few numbers of wide rings.

As electrons pass through the sample, some of the electrons undergo coherent and incoherent scattering, which makes it possible to implement two modes of imaging: bright field (BF) and dark field (DF). In the first mode, the aperture diaphragm of the objective lens (Figure 3) transmits the electrons of the incident beam. In this case, areas with a greater thickness or containing heavier chemical elements look darker due to the larger proportion of scattered electrons. Amorphous inclusions against the crystalline areas will also look darker due to a strong incoherent scattering. For polycrystalline thin films, the grains having different orientations will differ in brightness due to the difference in the fraction of electrons involved in the coherent (Bragg) scattering. In the DF mode, the image of the crystalline sample is built using the selected diffracted beams (with a DF aperture), while the transmitted beam does not participate in the image formation. This mode, in particular, allows visualization of a second phase inclusions and to obtain contrast images of crystalline inclusions in an amorphous material.

Scanning a sample by a very sharp (less than 0.1 nm in diameter) focused electron beam makes it possible to implement the High Angle Annular Dark Field (HAADF) image mode. For this mode, an image is formed by those electrons that have been deflected (as a result of incoherent scattering) at such a large angle that coherent scattering no longer makes a significant contribution to the signal. Then, the regions with a less perfect structure appear brighter compared to the surrounding regions.

Another image mode can be realized at high magnifications ($> 10^6$) – High Resolution Transmission Electron Microscopy (HRTEM). In this mode, the interference of the diffracted beams with the primary one creates periodic lines corresponding to atomic planes. Measuring the distances between the lines allows to find the interplanar distances and thus to identify the observed phases. The Fourier transform of the image often facilitate to create such reflections, which are not seen on the diffraction picture due to their low intensity. The HRTEM is very powerful method in revealing extended defects such as dislocations, stacking faults and twin boundaries.

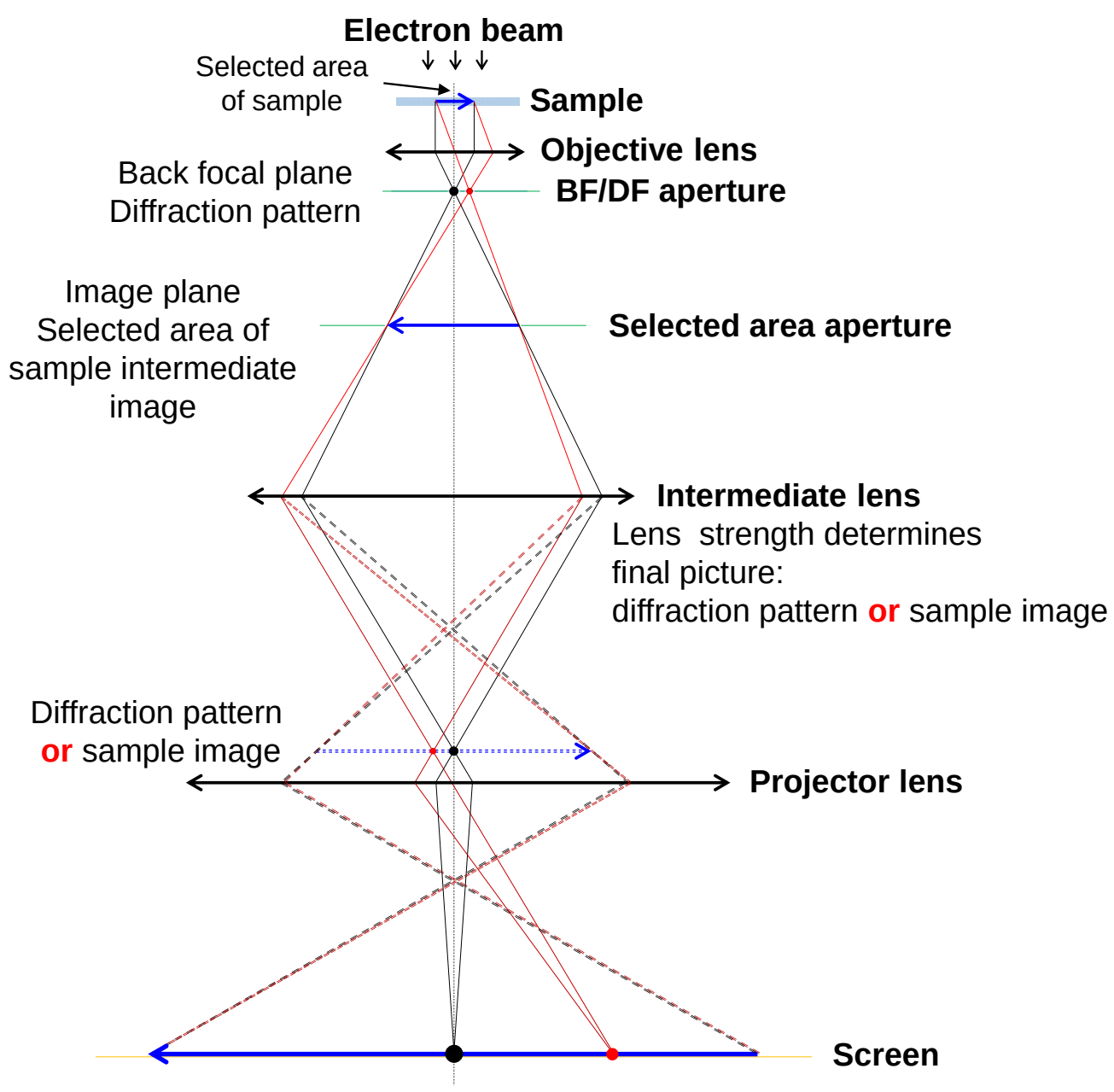

Figure 3. Diagram of the rays in two basic modes of TEM. Deflection angles in the figure are much larger than real ones (for better understanding). Only necessary lenses and apertures are shown; the real TEM has many more (that gives more flexible control). Intermediate lens changes the final picture on the screen (the ray's ways after this lens split on two: solid lines are for diffraction pattern; dotted lines are for image of the sample). *Bright filed / dark filed* (BF/DF) apertures are used for the corresponding mode choice. *Selected area aperture* is used to limit the area from which the diffraction pattern is obtained.

### *Scanning Electron Microscopy (SEM)*

SEM is based on registration of electrons outgoing from the sample surface during its irradiation by a scanning focused electron beam with a typical energy of several tens of keV. There are two main modes of SEM. In one of them, a current caused by secondary electrons (SE) is recorded. The SE are generated in the near-surface layer of the sample investigated due to the interaction of the beam with weakly bound electrons localized on the atomic shells and/or in the valence band of the semiconductor. The coefficient of SE emission depends on the local angle between the surface and the beam. Thus, by using beam scanning and registering the SE current, the surface morphology can be observed. Thereby, computer graphics allows creating its three-dimensional image.

In another SEM mode called Back-Scattering Electrons (BSE) [???], the yield of electrons elastically scattered from the sample surface is recorded. The BSE yield depends on the mass of the target atoms. This makes it possible to investigate by scanning the degree of compositional homogeneity of the sample surface layer.

In addition to SE and BSE modes, modern scanning electron microscopes are equipped with facilities for recording signals such as: Auger electron emission, characteristic secondary X-rays, electron beam induced current (EBIC), cathode-luminescence (CL), etc.

SEM can be used both in plan-view geometry and in cross-section of the samples. This, in particular, allows one to visualize and investigate the interval boundaries of various phases, including those formed by ion implantation.

*Energy Dispersive Spectroscopy (EDS)*

The method of EDS is based on the detection of characteristic X-Ray emission under an electron beam and is used in analytical electron microscopy. It allows exploring the elemental composition of near-surface layers of samples together with TEM and SEM images, surface morphology and other local characteristics. This method is especially useful for the identification of nano-objects observed by these techniques. The typical thickness of the probed layer is from several units to several tens of microns. EDS is often used to determine the distribution of elemental composition (elemental maps) over depth by probing the cross-section of a sample.

*X-ray Diffraction (XRD)*

To study the structure of implanted $Ga_2O_3$ layers, the X-ray Diffraction method is most often used. If a monochromatic X-ray beam is directed at an angle $\theta$ to some family of atomic planes, a diffracted beam occurs under the condition:

$$2d\sin\theta = n\lambda, \tag{1}$$

where $d$ is the interplanar distance, $\lambda$ is the wavelength, $n$ is an integer (the diffraction order). XRD allows to obtain several basic characteristics of the investigated samples as follows:

1) If the phase composition of the sample is not known in advance, then it can be identified using the values of $\theta$ at which the diffraction maxima are observed.
2) The method makes it possible to establish whether the analyzed layer of the sample has a single-crystalline, polycrystalline or amorphous structure. In the first case, in the set of diffraction peaks recorded with variation of the angle $\theta$, there are only those that correspond to different values of $n$ for the same family of atomic planes; in the second case, the peaks corresponding to different families of planes are observed; in the third case, usually there are no peaks at all.

3) The intensity and width of the peaks are used to judge the degree of structural perfection of single-crystalline layers. For this purpose, a High-Resolution X-Ray Diffraction (HRXRD) is used with an angular resolution of the order of a few angle seconds.

4) The value of the shift of the XRD peak positions versus the reference peak positions are used to determine the changes in the lattice parameters associated with the presence of the impurities and defects.

5) In HRXRD, the presence, character and magnitude of strain in single-crystalline layers, in particular those modified by ion irradiation, can be determined from the shape of the diffraction peaks.

In the case of using the XRD method to study layers modified by ion implantation, the analysis of data can be complicated when the thickness of the analyzed layer is significantly larger than the depth of ion penetration, as well as when a gradient of parameters over depth takes place. (The effect of this factor can be reduced by using small angles between the sample surface and the incident beam).

*X-ray Absorption Fine Structure*

This method includes two modes: X-ray Absorption Near Edge Spectroscopy (XANES) and Extended X-ray Absorption Fine Structure (EXAFS). In both modes, a monochromatic X-ray beam (from synchrotron radiation) with varied wavelength ($\lambda$) is directed onto a sample, and the spectra of X-ray absorption are studied. In XANES, a narrow energy region located below the position of the absorption edge of the atoms of a given material is studied, and in EXAFS, a wider energy region located above this edge is investigated. X-ray absorption is usually determined either from the values of current caused by the X-ray induced electron emission or from the intensity of X-ray fluorescence.

XANES spectra provide information on the chemical state of a definite atom, in particular about its oxidation state, effective charge and immediate environment, while the analysis of EXAFS data allows one to investigate the positions of atoms located not only in the immediate vicinity of the atom of interest, but also at greater distances, i.e., to determine the radial distribution functions.

*X-ray Photoelectron Spectroscopy (XPS)*

This method is based on the phenomenon of external photoeffect – the emission of electrons under the impact of characteristic X-ray radiation. Thereby, the kinetic energy ($E_{kin}$) spectrum of the emitted electrons is recorded with an energy analyzer. For the ideal case of absolutely pure elementary substance, the value of $E_{kin}$ is given by the relation:

$$E_{kin} = h\nu - E_{bind} - \varphi, \qquad (2)$$

where $h\nu$ is the energy of the incident photon, $\varphi$ is the difference between the work functions of the investigated substance and the material of the first electrode of analyzer, $E_{bind}$ is the "binding energy" defined as the energy required to remove an electron from a certain energy level of an atom or from the valence band of a dielectric or semiconductor to an infinite distance. If the substance contains atoms of different chemical elements, i.e. for a chemical compound or a solid solution, the spectral position of the corresponding line in the photoelectron spectrum undergoes the so-called chemical shift, which makes it possible to determine the chemical composition of the material. The spectra contain large amount of information, the detailed consideration of which is beyond the scope of this article and is described in the specialized literature (e.g.[37]).

Photoelectrons are emitted by atoms located not only directly on the surface, but also at a certain depth. Usually, the thickness of the analyzed layer is 2–5 nm. If there are some contaminants on the surface or a natural oxide, they are removed by sputtering with a low-energy (~ 1 keV) ion beam. Ion etching is used also to analyze the inhomogeneity of the chemical composition over depth.

When studying chemical compounds or solid solutions, the peaks in the XPS spectrum are often asymmetric, including the cases of overlapping of maxima corresponding to atoms of one and the same element in different chemical states. So, in $\beta$-$Ga_2O_3$ some gallium atoms are bonded to four neighboring oxygen atoms, while others – to six (tetrahedral and octahedral Ga configurations, respectively) (Figure 2). It is customary for oxides to use the term "oxidation state", meaning the number of oxygen atoms associated with a given one. To find the fraction of atoms in a particular oxidation state, the peak is decomposed into the functions that combine Gaussian and Lorentzian (deconvolution procedure). Interpretation of the resulting set of such functions is not always easy due to the ambiguity of the deconvolution procedure (this often requires some additional information or *ab initio* modeling). Another difficulty is that the asymmetric form of peak may be caused also by the inhomogeneity of phase composition in the analyzed layer. For example, in the case of $Ga_2O_3$, the phases of $\alpha$-$Ga_2O_3$ and $\beta$-$Ga_2O_3$ may be present, as well as the inclusions of elemental gallium and nonstoichiometric oxides

*Secondary Ion Mass Spectrometry (SIMS)*

This method is designed to reveal the chemical composition and incorporated species in the material studied. It allows measuring the distribution of impurity atoms over depth with a concentration sensitivity approaching $10^{13}$-$10^{14}$ $cm^{-3}$. SIMS is based on the phenomenon of sputtering of solids by an ion beam. Although most of the sputtered atoms are electrically neutral, some of them are emitted in a charged state (secondary ions). With the help of electric and magnetic fields, these ions are separated according to their mass to charge ratio. In modern time-of-flight (TOF) mass-analyzers, the separation is carried out according to their drift time in some space of the spectrometer.

The use of the focused probing ion beam in conjunction with the surface scanning makes it possible to determine the impurity concentration at different areas of the surface with a lateral resolution reaching 50 – 100 nm. To obtain depth concentration profiles, ion etching is performed either by the probe beam itself or with the help of a separate ion beam.

It should be noted that when using the SIMS technique, some difficulties are encountered, especially for high impurity concentration, when the sputtering coefficient may be concentration-depended, (e.g. see [38])

*Scanning Probe Microscopy*

Scanning Probe Microscopy (SPM) includes several methods or modes, the common feature of which is raster scanning of the sample surface with a sharpened probe, machine recording and processing of the data. The advantage of SPM is its relative simplicity (unlike TEM, it does not require a complex procedure for the sample preparation), a large amount of information received and very high maximal resolution. The two most commonly used SPM methods are Atomic Force Microscopy (AFM) and Scanning Tunneling Microscopy (STM).

AFM is most often used to study the surface morphology (although its capabilities are much higher). It is based on the fact that, at small distances, the force of interaction between atoms located on the surface of the sample and on the tip of the probe strongly depends on this distance. Precise movements of the probe (or sample) in the horizontal and vertical directions are carried out using piezomotors. The feedback arrangement, which includes a laser beam, a mirror fixed on the probe mount ("cantilever"), and a photodetector, ensures the maintenance of a given distance between the probe and the sample surface. With the use of analog-to-digital converter and computer graphics, a 3D image of the relief is created. The typical resolution is a few tenths of nm, and the lateral resolution is of the order of a few nm.

STM is based on the phenomenon of electron tunneling through a potential barrier in the presence of a vacuum gap between the sample and the tip of the probe. At given applied voltage between sample and the probe, the value of tunneling current depends both on the local electrophysical properties of the sample surface and on the size of the vacuum gap. The feedback system uses the deviation of the tunneling current. In the Surface Relief mode, the system tracks and digitizes the vertical movement of the probe during scanning.

There is an additional mode of Scanning Tunneling Spectroscopy (STS), which provides information on the local properties of the sample, such as the work function, the position of the Fermi level in the band gap, *I-V* and *C-V* characteristics, etc. The values of spatial resolution in STS could be as high as of atomic scale, especially in ultra-high vacuum. Such a high resolution is possible due to the exponential dependence of tunneling current on the size of the vacuum gap.

Scanning Probe Microscopy can be used (in combination with other methods) to study many processes during ion implantation: a change in conductivity of local arears due to the activation of implanted impurities, formation/annealing of disordered regions and defect clusters, synthesis of phase inclusions, etc.

*Rutherford Backscattering Spectrometry with Channeling (RBS/C)*

This method is based on measuring the energy spectrum of high-energy light ions (most often it is $He^+$ with $E$ ~ 1 MeV), after Rutherford scattering from a crystalline sample. A well-collimated ion beam falls on the sample surface, and the yield of ions scattered at angles close to 180° is recorded[39 40 41]. Let us consider the RBS/C application in radiation damage under ion irradiation study. In this case, the energy spectra of un-implanted samples are measured first (As a rule, crystals with a surface oriented perpendicular to the family of planes with low indices are used, since in this case the structural channels are more wide, so the effect of ion channeling is better expressed. The RBS spectrum is recorded with two orientations of the sample relative to the incident beam: with high misorientation (usually 7°), and with no misorientation (“random” and “channeling” spectra, respectively). In the first case, there is no channeling, and the spectrum coincides with the spectrum of the amorphous state. This spectrum has a steep drop on the high energy edge. For a clean surface, this edge corresponds to the loss of ion energy due to the scattering from atoms located immediately on the surface. In the second case (“channeling” spectra), most of the incident ions move along the channels only with seldom scattering and, in the spectrum of back-scattered ions, the yield from ions that have not entered the channels, as well as those that left the channeling mode due to non-ideal structural perfection of the crystal, plus the yield from ions scattered from defects blocking the channels, are dominating. For structurally perfect samples, the yield of ions back-scattered in the near-surface region is very small. For the crystals having less structurally perfect layers, the channeling spectrum is characterized by the presence of a region with an increased value of the yield. The energy width of such regions is defined by the thickness of the layer with a damaged structure, and the yield value characterizes the degree of damage. Spectra processing using a special software[39 40 41] allows determination the degree of damage as a function of depth. If an amorphous layer exists near the surface, then the yield of back-scattered ions from it is equal to the yield of random spectra.

Thus, the RBS/C method allows studying the processes of accumulation and annealing of damage, including the process of solid-phase recrystallization of amorphous layers. In addition, the existence of isolated disordered regions (clusters) or extended defects which cause local stresses and, as a consequence, distortion of atomic rows forming the channel “walls”, leads to increased yield from the layers located deeper than the implanted one. This allows revealing the presence of such objects.

In the case when implantation causes a transition of a substance to another crystalline phase (i.e., a polymorphic transformation), the most careful analysis of the RBS spectrum is required, since such a transformation can lead to a similar effect as that of accumulation of radiation defects or amorphization.

Another application of the RBS/C method is to study the concentration profiles of implanted impurities with atomic masses exceeding the atomic mass of the material studied. This is possible due to the fact that the energies of the ions scattered from such impurities lie in the region of higher energies in comparison with the energy of the ions scattered from the intrinsic target's atoms, so that the corresponding peak in the RBS spectrum is separated by a certain gap from the main spectrum of substance. In addition to the determination of concentration profiles for such impurities, the RBS method makes it possible to determine the fraction of impurity atoms located on the lattice sites. Indeed, if an impurity atom is on a lattice site, this atom turns out to be screened by intrinsic lattice atoms located in the same atomic row, while atoms located in random positions or interstitial ones are not screened and give contribution to the yield of backscattered ions.

Thus, the RBS method provides important insights into the degree and kinetics of disordering, the lattice positions of heavy impurities and their distributions, as well as into the kinetics of damage annealing. However, the sensitivity of the method is limited to a few percent of the total concentration of atoms. In addition, this method, as a rule, does not provide direct information on the defect’s types.

*Electron Beam Induced Current (EBIC)*

The method is based on the generation of electron-hole (*e-h*) pairs in a semiconductor by an electron beam. The *e-h* pairs are separated in the space-charge region of a *p-n* junction or a Schottky barrier and induce a current. Electron beam scanning permits to search a change of current over a surface. In particular, when the electron beam (in the SEM method) probes a region near the extended defect, the current is decreased due to an increased rate of *e-h* pairs recombination, and a respective contrast appears in the image. The same effect occurs in the cross-sectional mode of SEM. As it was demonstrated[42], the EBIC also allows determining the diffusion length of minority charge carriers.

*Deep Level Transient Spectroscopy*

Deep Level Transient Spectroscopy (DLTS) allows determination of energy levels located in the semiconductor band gap at large distances from the band edges [43]. The method is based on measuring the transient capacity of the Schottky barrier or *p-n* junction after the application of a voltage pulse providing non-equilibrium filling of deep energy levels by carriers. After the end of the pulse, the capacitance relaxes to its initial value due to the thermal ejection of the trapped charge carriers into the allowed band. The relaxation rate depends on temperature according to the Arrhenius law, which makes it possible to determine the spectrum of energy levels and their concentration by computer data processing.

One of the varieties of this method is the Optical Deep Level Transient Spectroscopy (ODLTS), when filling of deep levels by carriers is performed not by electric, but by optical pulse[44 45].

Another variation of this method is the Deep Level Optical Spectroscopy (DLOS), which differs from the conventional DLTS method in that after filling the levels not the temperature but the energy of incident photons is changed [21]. DLOS has the following advantage over the conventional DLTS: the levels can be determined across the entire band gap, while the traditional DLTS is applicable to deep levels only.

*Photoluminescence (PL) and Cathodoluminescence (CL) spectroscopy*

The method of PL spectroscopy is based on the analysis of photoluminescence spectrum arising from radiative electron transitions from higher to lower levels in the semiconductor band gap, as well as from interband transitions. For the PL excitation, usually monochromatic light sources (lasers) are used.

By using continuous spectrum light sources with monochromator the Photoluminescence Excitation (PLE) spectra can be measured, which provide some information about PL mechanisms. Pulsed excitation is used to determine the PL kinetics. It allows to measure the rate of nonradiative recombination associated with defects. Such information is important particularly for study the annealing processes of the damage accumulated under ion irradiation. The PL method is also used to determine the emission parameters of impurity centers.

In the method of Cathodoluminescence Spectroscopy, in contrast to the PL method, excitation is carried out by irradiating the sample with electrons. This method is especially often used to study the wide band gap semiconductors, for which the PL spectroscopy requires the use of sources of hard UV radiation. An important advantage of the CL spectroscopy is also the ability to estimate the depth (distance from the sample surface) at which the majority of certain luminescent centers are located. For this aim, the energy of incident electrons, and hence the thickness of probed layer, is varied. When studying the samples subjected to ion implantation, this variation makes it possible to distinguish the contributions to luminescence of intrinsic impurities or defects which are located uniformly over the sample depth, and that of extrinsic ones (introduced by ion implantation into the near-surface layer of a semiconductor).

*Electron Paramagnetic Resonance*

For semiconductors, the Electrical Paramagnetic Resonance (EPR) method provides important information about the nature (configuration, elemental composition) of impurities and defects and their behavior when external conditions change, e.g., when irradiated samples are annealed. This method is based on the Zeeman effect – the splitting of energy levels of paramagnetic centers in a magnetic field. Paramagnetic centers contain unpaired electrons with spin. When a magnetic field with induction $\boldsymbol{B}$ is applied, the spin-degenerate energy levels are split. The difference $\Delta E$ of energy levels with and without a field is proportional to $\boldsymbol{B}$ and the coefficient of proportionality "$g$" is called Lande factor $g$ ($g$-factor). In a crystal, the electric fields of neighboring atoms (ions) surrounding a given impurity atom or defect affect the value of the $g$-factor. As the intracrystalline field is generally anisotropic, the $g$-factor is a tensor.

When measuring the EPR spectrum, in addition to the constant magnetic field, an ultrahigh-frequency electromagnetic field with a frequency $\boldsymbol{\nu}$ is applied to the sample. This field undergoes a resonance absorption when the value of its quantum $h\nu$ of energy coincides with the value of $\Delta E$ (In real situations, the resonance is usually achieved by varying the magnetic field $\boldsymbol{B}$ at a fixed value of $\boldsymbol{\nu}$).

Thus, by measuring the EPR spectrum and performing calculations in accordance with certain models, one can get insights into the structure of paramagnetic defect centers, their orientation with respect to the crystal axes. The paramagnetic defect centers concentration is determined by using reference samples with a known number of centers.

*Hall Effect measurements*

When current flows through the sample, the Lorentz force acts on the movement of charge carriers in a magnetic field, deflecting their trajectories and resulting in an additional electric field. This phenomenon is named Hall Effect, which, together with the measurement of electrical conductivity, is used to determine type of conductivity, concentration and mobility of charge carriers in semiconductors. In the case when the analyzed layer is located on an insulating substrate, the layer concentration of charge carriers $n_s$ is determined from the relation connecting this parameter with the voltage between the measuring contacts, current, magnetic field and geometry of contacts. The average value of charge carrier's mobility μ is determined from the relation $\sigma_s = q\, n_s \mu$, where $\sigma_s$ is the layer conductivity, and $q$ is the charge of electron. The type of conductivity can be found from the sign of the Hall voltage for a certain direction of the magnetic field. The ionization energies of impurities or defect centers are found from the temperature dependence of the charge carriers' concentration. To determine the depth distribution of specific conductivity and charge carrier's concentration, the $\sigma_s$ and $n_s$ measurements should be provided in combination with the step-by-step etching.

# IV. PROPERTIES OF IMPLANTED $Ga_2O_3$ LAYERS

## A. Properties of impurities in β-$Ga_2O_3$ introduced by ion implantation

Ion implantation is a well-stablished technique, widely used for the introduction of impurities and defects in solid materials since early 1960's. In this technique, positive or negative ions, extracted from solid or gaseous sources, are accelerated through a potential difference and attain high kinetic energies. The accelerated ions usually pass through an electromagnet to select the desired kind of the ions and are directed to the target (sample). Colliding with the atoms of the sample, the ions transfer part of their kinetic energy to them, displacing them from the lattice sites. Displaced atoms (recoils), acquiring sufficient kinetic energy, displace other atoms, and so on. Vacancies remain in place of the atoms displaced from the sites. This process leads to the formation of radiation defects (discussed in more detail in Section IV B). Here we are interested in the fate of the implanted ion, slowing down to an energy at which the ion is no longer able to produce recoils, enters the next vacancy it creates, becoming a substitutional atom, or stops at the nearest interstitial site, becoming an interstitial atom.

This process makes it possible to introduce into the semiconductor ions of any impurity ("ion" doping), including donors and acceptors. However, due to the presence of radiation defects, the donor and acceptor properties of the introduced impurities usually appear only after annealing at sufficiently high temperatures, which eliminate defects or reduce their concentration. Annealing is carried out most often either in special furnaces for a long time from several minutes to several hours ("furnace annealing", FA), or by lamp heating for several seconds ("rapid thermal annealing", RTA), less often - by other methods (current pulse, pulsed laser irradiation, etc.).

Nowadays, countless research groups around the world make regular use of this technique for modification of all sorts of materials, including semiconductors, metals and insulators, in the form of bulk, thin films, nanostructures. Ion implantation is the key method of modern microelectronic technology.

Among many others impurities introduced into β-$Ga_2O_3$ by ion implantation can be grouped into three main categories in accordance with the application of $Ga_2O_3$. The first category includes donor and acceptor impurities used for increase or decrease the majority charge carriers concentration – electrons when creating Ohmic contacts, forming MOSFET channels, producing high-resistance regions to reduce the edge leakage currents and increase breakdown voltages, etc. The second category includes impurities of rare earth elements and some transition metals used to create light-emitting centers. The third category includes light gaseous species – hydrogen, deuterium and helium.

The most often used impurities of the first category are the tetravalent elements of the 4$^{th}$ group of the periodic table - Si, Ge and Sn. They replace the trivalent gallium atoms in gallium oxide and create shallow donor levels.

The electrical activation of Si implanted into an unintentionally doped (UID) float zone (FZ) β-$Ga_2O_3$ crystals with surface orientation (010) was studied in Ref. [46]. Using multiple-energy implantation of $Si^+$ with the maximum ion energy $E$ of 120 keV, a 150 nm deep doped layer was formed having a box-like depth profile with Si concentrations in the range of $1\times10^{19}$ – $1\times10^{20}$ $cm^{-3}$. Post-implantation isochronous annealing was carried out in a nitrogen atmosphere at temperatures $T_{ann}$ = 700 °C – 1100 °C with a step of 100 °C (30 min at each $T_{ann}$). Depth distributions of electron concentration were determined by electrochemical capacitance-voltage (ECV) measurements. The Si concentration profiles were also measured by the SIMS method. It was shown that, up to $T_{ann}$ = 1000 °C the Si profiles are almost the same as in the as-implanted sample. At $T_{ann}$ = 1100 °C, the profile becomes nonmonotonous: near $R_p$ (mean projected range of ions), there is a sharp increase in concentration, and in the region close to the surface, a similarly sharp decrease takes place. It is assumed that the Si redistribution is a consequence of the impurity diffusion to the region of maximum damage (i.e., segregation due to the interaction of impurities and defects).

Figure 4 shows the dependencies of estimated (by ECV measurements of $N_d – N_a$) concentrations of electrons in Si-implanted β-$Ga_2O_3$. The activation of the implanted impurity begins at 800 °C and reaches a maximum value at 1000 °C, decreasing with a further increase in $T_{ann}$, except for the case of the maximum Si dose, for which the dependence of electron concentration on $T_{ann}$ is monotonous, but the fraction of activated impurity is very small (does not exceed 10%). The decrease in electrical activity of the impurity at high doses is apparently due to an increase in the fraction of Si atoms forming complexes with unannealed radiation defects (the concentration of the latter increases with the dose). The nonmonotonous behavior of the concentration of charge carriers with $T_{ann}$ is possibly related to the out-diffusion of Si. However, the refinement of the mechanism of this impurity behavior requires further investigations.

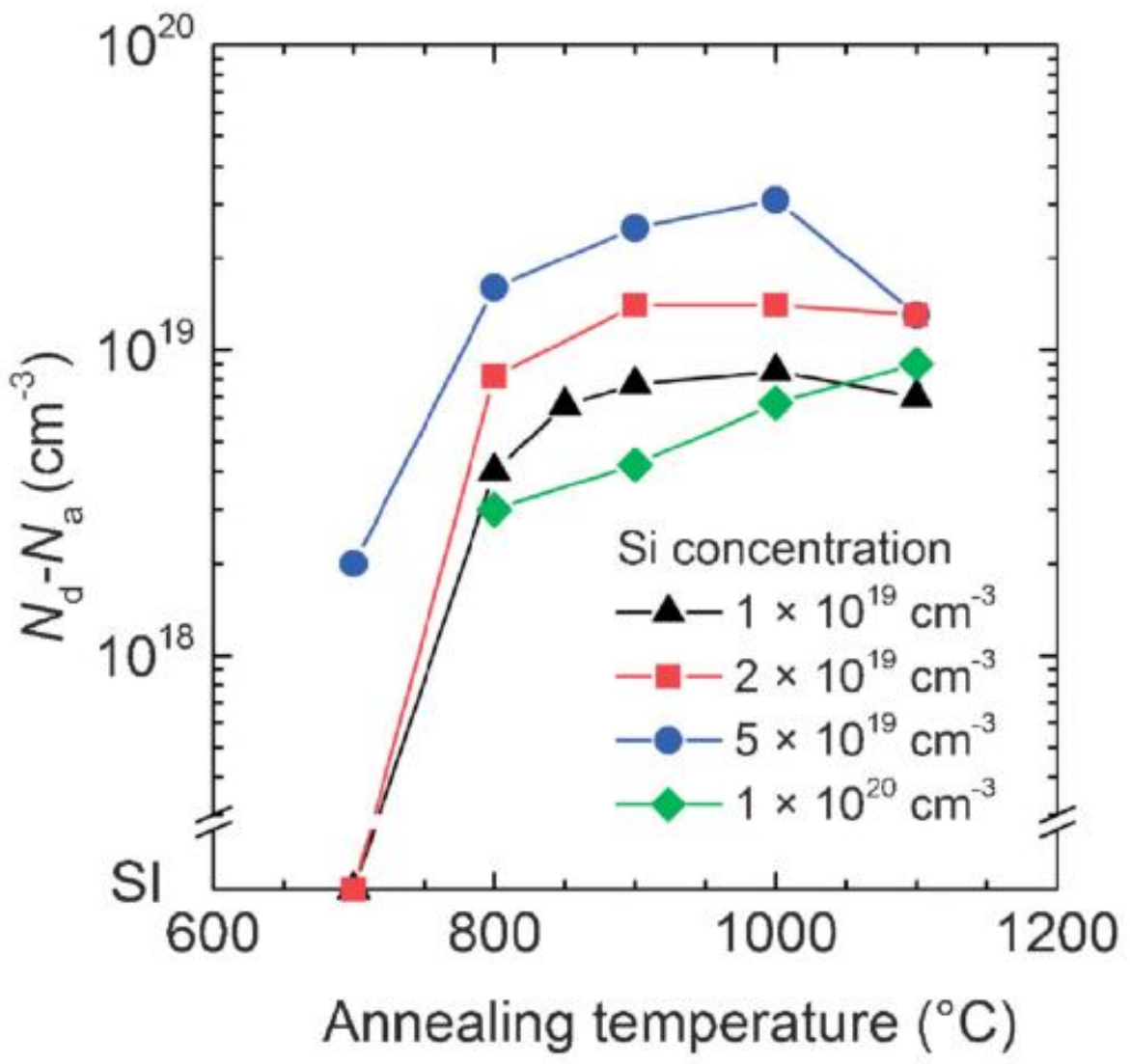

Figure 4. Annealing temperature dependencies of electron concentration in the Si-implanted UID β-$Ga_2O_3$ layers. Reprinted with permission from [46] Copyright (2013) The Japan Society of Applied Physics

The diffusion behavior of silicon implanted in β-$Ga_2O_3$ during rapid thermal annealing (RTA) ($T_{ann}$ = 1100 °C, annealing time of 10 – 120 s) was studied by the SIMS method[47]. As in [46], the Si box-like profile was created by multi-energy implantation (maximum ion energy $E$ = 120 keV). Unlike Ref. [46], the surface orientation of the samples was (-201). It was found out that the annealing atmosphere had a significant effect on the Si diffusion: upon annealing in oxygen, the redistribution of impurities was much more pronounced than upon annealing in nitrogen. Moreover, in the case of oxygen annealing, there was a significant loss (out-diffusion) of silicon, which was not observed during annealing in nitrogen. According to the model proposed by the authors, this difference is due to the fact that annealing in nitrogen creates an increased concentration of gallium vacancies, which capture interstitial Si atoms and thereby slow down their diffusion.

In Ref. [48] of the same research group, the Si diffusion at 1150 °C was calculated on the basis of a simple (phenomenological) model that takes into account the concentration dependence of the diffusion coefficient. By comparing the experimental profile with the calculated one for different doses, the "low-concentration" diffusion coefficient at 1150 °C and the coefficient characterizing the concentration dependence of out-diffusion were determined. However, the use of only one $T_{ann}$ for comparison and the phenomenological nature of the model don’t allow judging on the applicability of the results obtained for other experimental conditions. In addition, the results may depend on the surface orientation due to the anisotropic structure of the monoclinic modification (β-phase) of $Ga_2O_3$.

Similar results were reported by this group for the implantation of two other (heavy) donor impurities of the fourth group – $Ge^+$ and $Sn^+$ [49 48] and RTA with $T_{ann}$ = 1100 °C. In the case of as-implanted samples, unlike the case of $Si^+$ ions, extended end-of-range (EOR) defects were revealed by the cross-sectional TEM method. The authors discuss the difference in the diffusion mobility of Ge and Sn in terms of the difference in atomic radii, which is responsible for the local strains arising in the lattice around the substitutional atoms, and come to the conclusion that there is no correlation: the observed faster diffusion of Ge compared to Sn and Si cannot be explained by the difference of atomic radii of these impurities.

The diffusion behavior of implanted acceptor impurities like nitrogen and magnesium was studied by SIMS[50]. $Mg^+$ was implanted into a β-$Ga_2O_3$ (001) substrate doped with Sn at a concentration of $2\times10^{18}$ $cm^{-3}$, and $N^+$ – into a UID β-$Ga_2O_3$ (001) substrate with a charge carrier concentration of $2\times10^{17}$ $cm^{-3}$. The maximum concentration of magnesium in the implanted samples was ~ $10^{19}$ $cm^{-3}$, and of nitrogen ~ $10^{18}$ $cm^{-3}$. The $R_p$ values for $Mg^+$ and $Si^+$ ions were approximately the same (~ 0.5 μm). Post-implantation annealing was carried out for 30 min at $T_{ann} \geq 600$ °C with a step of 100 °C in a nitrogen atmosphere.

It was found out that the redistribution of magnesium began at $T_{ann}$ = 900 °C, while a significant redistribution of nitrogen occurred only at $T_{ann}$ as high as 1100 °C. This circumstance is important, since it allows, when using nitrogen ions, to increase the annealing temperature to higher values in order to provide more complete activation of the implanted impurity.

Interestingly, on the concentration profiles of Mg and N reported in [50], the regions exist where the concentration of these impurities practically coincides with the concentration of the impurity (silicon) in the initial samples. This indicates the formation of complexes consisting of the implanted and the original impurities due to donor-acceptor pairing: ($Mg^-$ – $Si^+$), ($Mg^-$ – $Sn^+$) and ($N^-$ – $Si^+$). A similar phenomenon was observed for ion doping of Si and GaAs[51 52 53 54]. However, the effect of complex formation on the diffusion of impurities in β-$Ga_2O_3$ has not been studied. Thus, the physical nature of the difference in the diffusion behavior of Mg and N is not entirely clear. The difference may be due to the greater degree of radiation damage during the implantation of heavier ions (Mg), which was experimentally established in [50] using XRD. However, theoretical calculations performed by the DFT method [30] showed that, even in the absence of implantation, the effective activation energy of Mg diffusion is lower than that for N. This may be explained [30] by the fact that nitrogen diffusion is dominantly assisted by oxygen vacancies, while magnesium diffusion is assisted by Ga interstitials.

In contrast to shallow donor impurities, acceptor impurities in β-$Ga_2O_3$ introduce deep levels into the bandgap [23 29]. Therefore, their activation during annealing and, accordingly, the compensation of donors leads to an increase in the resistivity, which was experimentally confirmed by in situ doping during the β-$Ga_2O_3$ growth process[55 56].

In the case of ion implantation, defects such as gallium vacancies can also have a compensating effect [15 10]. In the work [57], nitrogen ions were implanted into epitaxial β-$Ga_2O_3$ layers with an initial electron concentration of $2\times10^{18}$ $cm^{-3}$, and a box-like profile of generated vacancies with a concentration of ~ $10^{21}$ $cm^{-3}$ was created. Measurements of the layer conductivity and the Hall effect of the $N^+$ irradiated samples demonstrate formation of a semi-insulating layer, the resistivity of which remains practically unchanged up to $T_{ann}$ = 600 °C. An increase in $T_{ann}$ to 800 °C causes the recovery of conductivity almost to its original level. This behavior indicates that the compensation of conductivity was caused not by the acceptor properties of nitrogen, but by the radiation defects. However, in the work[50], by measuring the leakage currents of the Schottky barrier it was revealed that at $T_{ann} \geq 800$ °C the resistivity of the nitrogen-implanted layer increases with the annealing temperature.

In Ref.[58], the compensating effect of implanted nitrogen in β-$Ga_2O_3$ (with a silicon concentration of $3.9\times10^{17}$ $cm^{-3}$) after annealing at 900 °C was established by measuring the *C-V* characteristics of the Schottky barrier. However, despite the fact that the concentration of implanted nitrogen was as high as $10^{20}$ $cm^{-3}$, the electron concentration decreased as a result of $N^+$ implantation only to the level of $5\times10^{16}$ $cm^{-3}$. This apparently was due to the low degree of nitrogen activation at such $T_{ann}$ and indicates the need to use higher annealing temperatures. All these experiments prove that the mechanism of compensation by $N^+$ implantation depends on the nitrogen dose and annealing conditions.

In the case of $Mg^+$ implantation, the behavior of the leakage current with annealing temperature was nonmonotonous[50]: the leakage current increased in the range 600 °C ≤ $T_{ann}$ ≤ 900 °C and decreased at $T_{ann} \geq 1000$ °C. Note, that in this publication for magnesium and in the work [58] for nitrogen, the leakage current after annealing became higher than that before annealing (when the compensation was caused apparently not by the impurity Mg or N atoms, but by the radiation defects).

Let's now consider the implantation of impurities belonging to the second group, i.e., the rare earth elements Eu, Er, Gd, and transition metals. The behavior of these impurities was studied dominantly in connection with their luminescence properties[59 60 61 62 63]. Implantation was carried out into bulk β-$Ga_2O_3$ and nanowires of β-$Ga_2O_3$. For $Eu^+$ implanted into bulk β-$Ga_2O_3$[60], the correlation between the implantation temperature, the temperature of post-implantation annealing, the Eu concentration in the +3 charge state, on the one hand, and the luminescence spectra and intensity, on the other hand, were investigated. It was demonstrated that the radiation defects played a key role in the luminescence properties and charge state of Eu.

In Ref. [64], the features of Raman scattering spectra and CL of β-$Ga_2O_3$ nanowires and bulk crystals implanted with $Cr^+$ and $Mn^+$ ions with $E$ = 150 keV and $D = 10^{15}$ $cm^{-2}$ were studied. The measurement of Raman peaks linewidth showed that the improving the structure during rapid thermal annealing of ion-irradiated nanowires began at 700 °C and finished at 1000 °C. The behavior of $Cr^+$-related cathodoluminescence at annealing was correlated with the structure improvement and it was the same for nanowires and bulk samples, while in the case of $Mn^+$ implantation, the behavior in these objects was different. The physical reasons for the similarity or differences in the impurity's properties for implantation in bulk samples and in nano-objects are not yet clear.

Among light gas ions (third category of implanted impurities), the most important is hydrogen[3]. It can penetrate as an unintended impurity during the growth of both films and bulk crystals. It is believed[3] that one of the main reasons of the commonly observed *n*-type conductivity in UID β-$Ga_2O_3$ may be the presence of residual hydrogen in the growth ambient. The donor properties of hydrogen were supported by theoretical calculations in the work[17]. It can also passivate deep acceptor impurity centers, reducing their efficiency [25]. However, it should be taken into account that hydrogen is a rapidly diffusing impurity and can escape from the semiconductor even at relatively low annealing temperatures (see below).

Interest to the hydrogen impurity has been especially increased after the exciting report[65], where the authors claim that they had fabricated a low-resistance β-$Ga_2O_3$ of *p*-type by diffusion of this impurity.

To study the diffusion behavior of hydrogen it is convenient to use its isotope deuterium (D), because deuterium is easier to be detected than H. The relationship between the diffusion coefficients of H and D is defined by the relation $D_H/D_D = (M_D/M_H)^{1/2}$, where $M$ is the mass of the corresponding atom[66]. The diffusion of deuterium introduced into β-$Ga_2O_3$ (-201) by ion implantation was studied in[67] [68]. Deuterium concentration profiles were measured by SIMS before annealing and after isochronous annealing (5 min each) at $T_{ann}$ = 450 °C – 650 °C. Comparison with the case of doping by plasma exposure (i.e., without ion implantation), allowed revealing the role of trapping hydrogen (or deuterium) by radiation defects (vacancies) in the diffusion mechanism. By comparing the experimental profiles with the calculated ones the activation energies of various processes involved in H and D diffusion were determined[68].

In Ref. [69], the structural and vibration properties of O-H centers in β-$Ga_2O_3$ were investigated by infrared polarized absorption spectroscopy. The hydrogen and deuterium were introduced into β-$Ga_2O_3$ (-201) samples by two ways: (a) by implantation and post-annealing up to 400 °C; (b) without implantation – by annealing in $H_2$ and $D_2$ atmospheres. Infrared absorption spectra were measured at two different light polarizations. By comparing the spectra for the cases of doping with only hydrogen, only deuterium, as well as with hydrogen and deuterium together and using the results of DFT calculations, the authors of [69] concluded that the dominant O-H center in $Ga_2O_3$ hydrogenated by annealing in a hydrogen-containing atmosphere or by implantation was a relaxed $V_{Ga}$-2H complex. In the work[67 68], it was shown that hydrogen introduced in $Ga_2O_3$ without ion implantation undergoes out-diffusion at lower temperatures than the implanted hydrogen. This difference may be caused by trapping of hydrogen by the radiation defects generated by ion implantation.

The implantation of light gas impurities ($H^+$, $He^+$) was used for the exfoliation of thin $Ga_2O_3$ layers[70] (as it will be pointed below, this technique is needed to solve the problem of the low thermal conductivity of $Ga_2O_3$).

In Ref.[71], the formation of bubbles in β-$Ga_2O_3$ (010) upon implantation of $He^+$ ions with $E$ = 160 keV and $D = 5\times10^{16}$ cm$^{-2}$, followed by annealing at 200 °C and then at 500 °C was studied using XRD, TEM, and AFM methods. The growth of bubbles known as blistering and the formation of cracks were the main processes leading to the exfoliation of implanted $Ga_2O_3$ layer.

Summarizing the existing knowledge, accumulated in physical aspect of ion doping of β-$Ga_2O_3$, it is obvious that it is still at the initial stage of development. In particular, the role of radiation defects in the activation of implanted impurities during annealing should be more thoroughly studied. What is most important in terms of practical application, it is to gain a deeper insight into the properties of impurities introduced by ion implantation in $Ga_2O_3$ samples fabricated by different methods.

## *B. Ion-induced damage in $Ga_2O_3$*

The generation of damage is one of the most important features of the ion implantation method. This can be used in many cases as a tool for materials modification, while in other cases the damage can have a detrimental impact on some of the material's properties.

When energetic ions penetrate through a substance, they transfer their energy to electrons and nuclei of the target. The type of ion-substance interaction depends on the ion energy and on the mass of the target atoms and impinging ions. The energy lost by the ions can be transferred to the nucleus (so called nuclear stopping power ($S_n$)), to the electrons (called electronic stopping power ($S_e$) and to the phonons (thermal loss). When the energy transferred to the atomic nucleus by incident ion exceeds a certain *threshold* value ($E_d$), a recoil atom and a vacancy are formed. The energetic recoils in turn can produce additional recoils and vacancies, and so on. As a result, each incident ion creates one or more groups of point defects, consisting of interstitial atoms and vacancies. Such groups of defects are called "displacement cascades".

In Table 2, the calculated parameters which characterize the interaction of three types of ions – $H^+$, $Ga^+$, and $Au^+$ ("light", "intermediate" and "heavy", respectively) with different energies are presented as examples. Calculations were carried out using the SRIM code [72].

Table 2. The parameters characterizing the interaction of ions with β-$Ga_2O_3$ target.

| Ion type | Energy (MeV) | Range (μm) | Straggling (μm) | Numbers of vacancies in the 100 nm layer per ion | |
|---|---|---|---|---|---|
| | | | | $V_{Ga}$ | $V_O$ |
| H | 0.01 | 0.09 | 0.04 | 255 | 258 |
| | 0.1 | 0.59 | 0.09 | 34 | 27 |
| | 1 | 8.74 | 0.37 | 4 | 3 |
| | 10 | $3.34\times10^{2}$ | 8.06 | 0 | 0 |
| | 100 | $1.89\times10^{4}$ | $3.76\times10^{2}$ | 0 | 0 |
| Ga | 0.01 | 0.01 | $0.03\times10^{-1}$ | $8.03\times10^{3}$ | $9.10\times10^{3}$ |
| | 0.1 | 0.04 | 0.02 | $6.89\times10^{4}$ | $7.82\times10^{4}$ |
| | 1 | 0.39 | 0.12 | $6.06\times10^{4}$ | $6.83\times10^{4}$ |
| | 10 | 2.63 | 0.33 | $1.12\times10^{4}$ | $1.22\times10^{4}$ |
| | 100 | 9.62 | 0.42 | $1.64\times10^{3}$ | $1.71\times10^{3}$ |
| Au | 0.01 | 0.01 | $0.02\times10^{-1}$ | $8.27\times10^{3}$ | $9.38\times10^{3}$ |
| | 0.1 | 0.02 | 0.01 | $7.13\times10^{4}$ | $8.08\times10^{4}$ |
| | 1 | 0.13 | 0.04 | $2.99\times10^{5}$ | $3.36\times10^{5}$ |
| | 10 | 1.38 | 0.27 | $8.57\times10^{4}$ | $9.67\times10^{4}$ |
| | 100 | 8.44 | 0.66 | $1.76\times10^{4}$ | $1.93\times10^{4}$ |

** The «straggling» means a root-mean-square scatter of ranges. The range and straggling refer to the semi-infinite target; vacancies of Ga ($V_{Ga}$) and O ($V_O$), their numbers in the 100 nm layer are indicated.*

The influence of mass and energy of ions on the type of ion trajectory and on the defect formation for the same examples is illustrated in Figure 5.

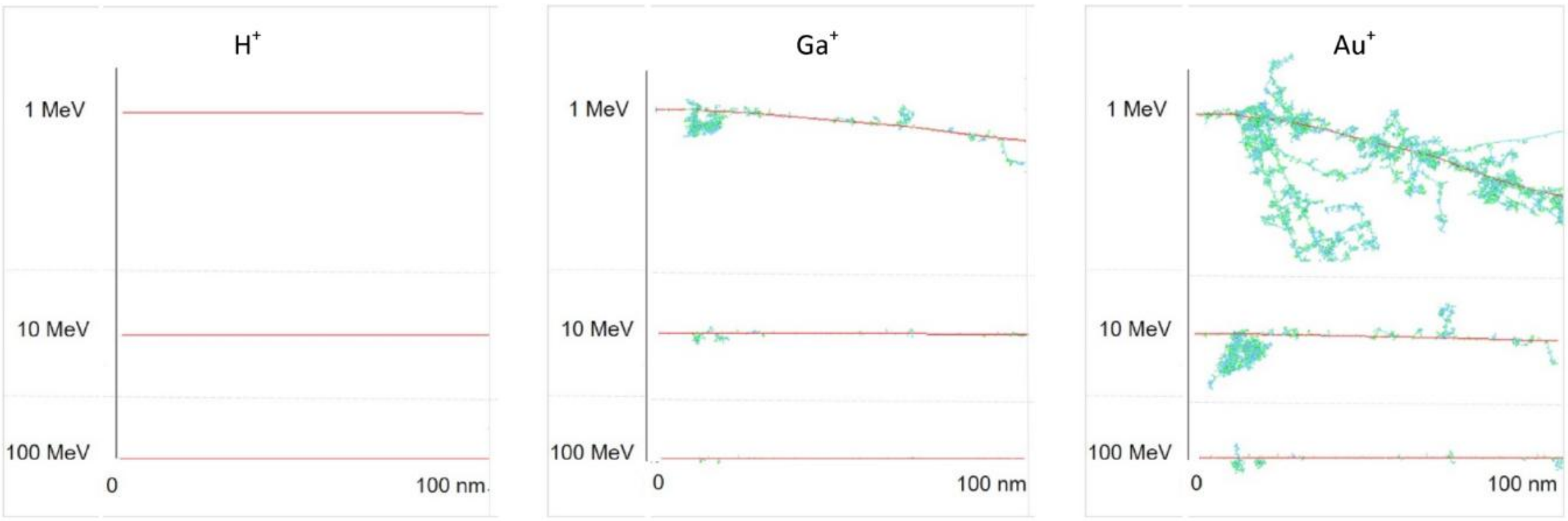

Figure 5. SRIM simulations of damage induced in β-$Ga_2O_3$ by irradiation with $H^+$, $Ga^+$, $Au^+$ ions at different energies. The ion trajectories are shown by the red lines while the Ga and O atoms displaced from their lattice positions (recoils) are shown in green and blue, respectively.

Figure 5 clearly shows the drastic differences between the damage induced by irradiation with ions of different masses and energy. While the protons with the indicated energies penetrate through the subsurface layer (~ 100 nm thick) practically without producing recoil cascades and with very small nuclear energy loss (hence, with a negligible damage), the heavy ion (here it is $Au^+$) with typical energy $E \leq 1$ MeV produces large cascades near the surface. With an increase in the energy of heavy ions, the value of $S_n$ decreases, while $S_e$ increases, so the degree of damage in the ~ 100 nm subsurface layer decreases. (However, at very high energies (not shown in Figure 5) of so-called "swift ions", the role of electron losses becomes dominant in damage). The behavior of ions with intermediate masses (e.g., $Ga^+$) is similar to the behavior of $Au^+$ ions with higher energy, as seen in Figure. 5.

Experimental investigations of damage in β-$Ga_2O_3$ under ion irradiation (excluding the case of swift ions) can be conditionally divided into three types. The investigations of the first type are devoted to the study of the damage degree and its distribution over depth, as well as to the study of the structure recovery caused by thermal post-irradiation annealing. The main research method used in these works is the RBS/C. The second type of investigations concerns mainly the point radiation defects – vacancies, intrinsic interstitials, and simple complexes. The main methods for identifying such defects are the luminescence spectroscopy and DLTS (with their varieties). These methods make it possible to determine the energy levels of point defects and to compare them with the results of theoretical modeling. Such studies are possible mainly under the conditions of a sufficiently low level of damage, which occurs typically for low doses of irradiation with relatively light ions.

In the studies of the third type, the main attention is paid to the behavior of implanted impurities, while the damage is of interest mainly insofar as it affects the activation of implanted impurities. These studies were considered in the section IV A. The works devoted to irradiation by swift ions were considered, in brief, at the end of section IV B.

Let us first consider the studies of the first type. In Ref. [73], the RBS/C method was used to study the dose dependence of the damage degree in β-$Ga_2O_3$ upon irradiation with $P^+$, $Ar^+$, and $Sn^+$ ions. Single crystalline bulk β-$Ga_2O_3$ (010) was used. Energies (240 – 400 keV) were chosen so that the thickness of the implanted layer was the same for different ions – about 250 nm; the doses $D$ were varied in the range of $1\times10^{11}$ – $2\times10^{15}$ $cm^{-2}$. To minimize the channeling effect during implantation, the ion beam was 7° off axis. It was found that the depth distributions of damage approximately corresponded to those calculated using the SRIM code[72]. In contrast to the behavior typical of semiconductors such as Si, Ge, and GaAs, the damage concentration did not reach 100% (the level of amorphization) with increasing dose, but reached a saturation value of ~ 90% (see Figure 6).

A similar behavior was previously observed in some other wide band gap semiconductors with a high degree of ionicity, such as ZnO and GaN [74] [75]. For β-$Ga_2O_3$, the degree of ionicity is lower, than that of ZnO and GaN, and the saturation level is higher. In addition, for ZnO and GaN, the dose dependence of damage was characterized by the presence of two plateaus (intermediate and final), while for $Ga_2O_3$ the intermediate plateau was less pronounced and not in all cases (e.g., it was absent for $Ar^+$). The authors interpret the data obtained for the dose dependence of damage employing the model considered in [76]. The latter is based on the assumption that upon ion irradiation, two types of primary defects are simultaneously generated: (a) elementary – single Frenkel pairs, which at each stage of irradiation partially recombine with defects generated at the previous stages, and (b) – clusters. If the clusters are not amorphous, the model predicts the saturation of damage at a level less than 100%. The authors of [73] concluded that for a correct interpretation of the RBS/C data on the distribution of damage over depth, it is necessary to take into account also the bending of the channels due to the strains produced by defect clusters. It would be interesting to apply for the interpretation of the experimental results reported in [73] the models previously used in studies of damage under ion irradiation of other wide band gap semiconductors, e.g. GaN and ZnO[77] [78] [79] [80] [81].

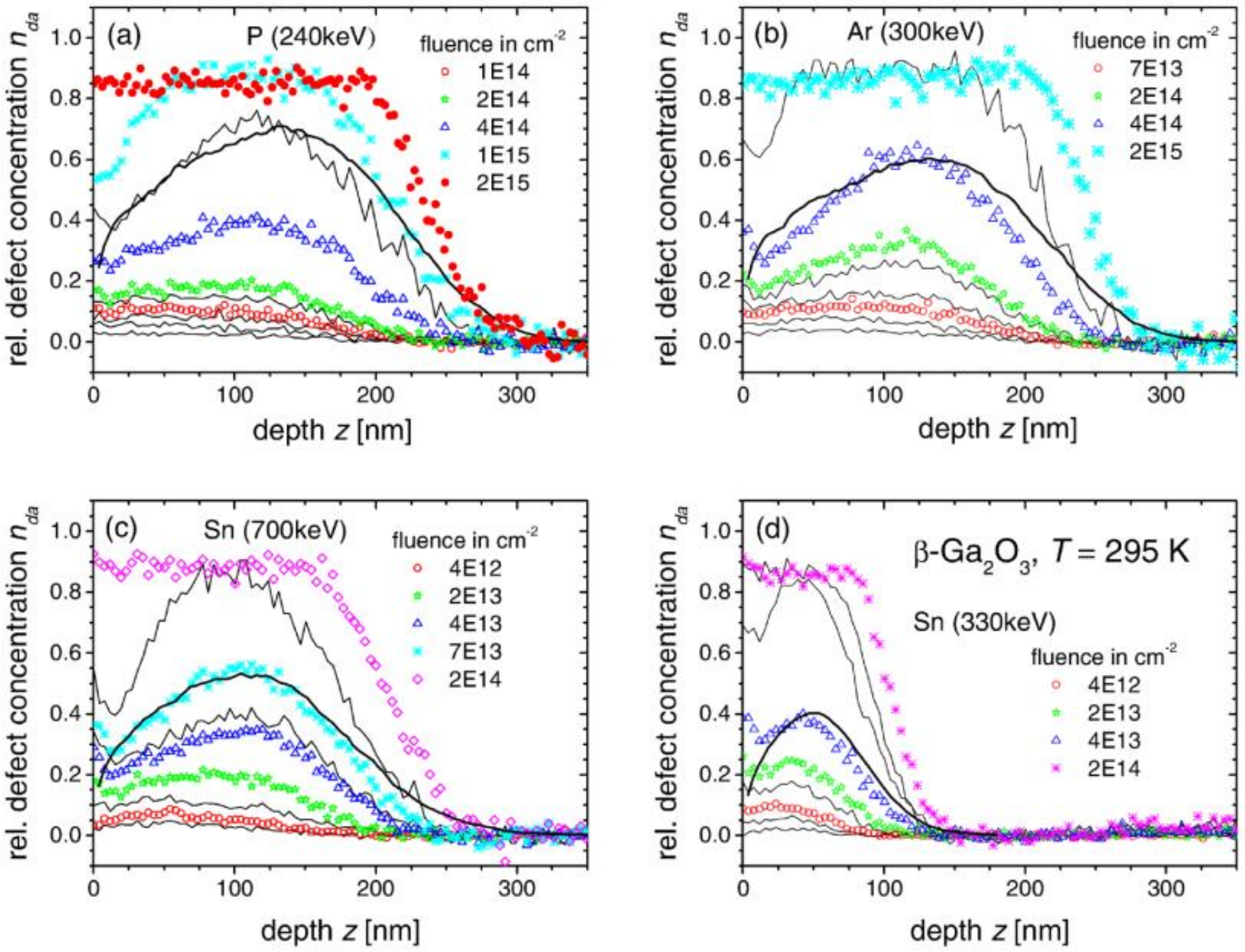

Figure 6. Relative damage level $n_{da}$ versus depth $z$ for β-$Ga_2O_3$ implanted at room temperature with (a) 240 keV P, (b) 300 keV Ar, (c) 700 keV Sn and (d) 330 keV Sn ions. For clarity only selected ion fluences are indicated. The thick solid line is the depth distribution of primary displacements as calculated by SRIM given in arbitrary units for comparison. Reprinted with permission from [73] Copyright (2016) Elsevier B.V.

The RBS/C method was also used in [59] to study the damage in β-$Ga_2O_3$ with (-201) surface orientation upon ion implantation of the rare earth element Eu (atomic mass $M$ = 152). These results are somewhat different compared to [73]. The difference, as noted in the article[73], is maybe due to both factors: a different surface orientation and a difference in the thickness of implanted layer. These factors are important when the surface serves as a drain for mobile defects. The energy of $Eu^+$ ion utilized was 300 keV, the doses ($D$) were $1\times10^{13}$, $1\times10^{14}$, $1\times10^{15}$ and $4\times10^{15}$ $cm^{-2}$. The mean projected range $R_p$ of $Eu^+$ was about 50 nm. It was found that, at the minimal dose used, the distribution of damage practically coincided with the vacancy distribution calculated using the SRIM code[72], but it differs significantly at higher doses (illustrated in Figure 7). For $D = 1\times10^{14}$ $cm^{-2}$, the damage decreases monotonically with depth and, for $D = 1\times10^{15}$ $cm^{-2}$, the distribution acquires a two-peak shape: the first peak is adjacent to the surface, and the second one (bulk peak) is located at a depth exceeding $R_p$. Moreover, in the area of the first peak, in contrast to the data[73], the degree of damage reaches the amorphization level.

With a further increase in the dose to $4\times10^{15}$ $cm^{-2}$, the near-surface peak expands, then it merges with the bulk one, and the total layer thickness with the level of damage corresponding to amorphization reaches a value close to $R_p$.

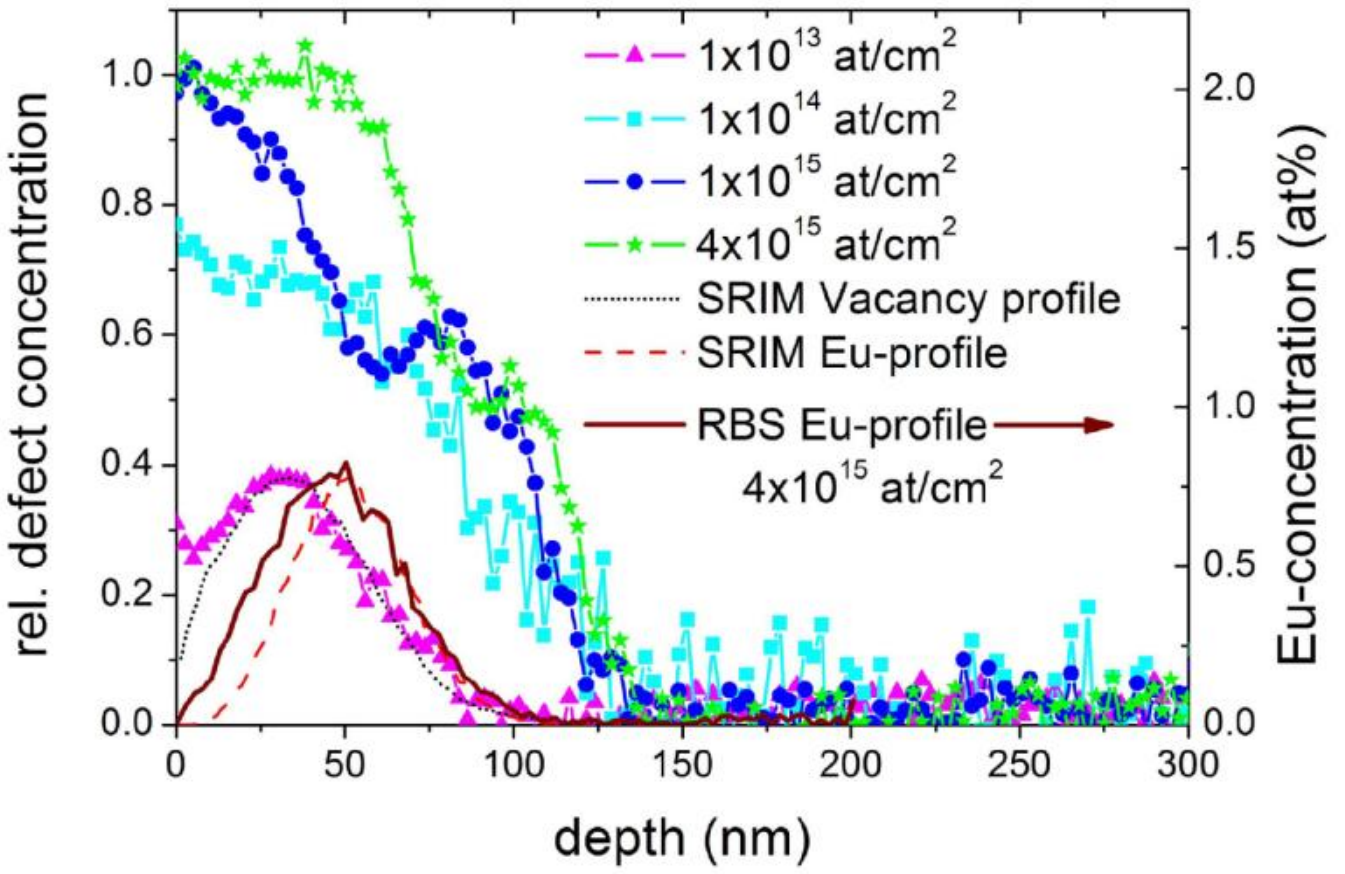

Figure 7. The relative defect concentration as a function of depth for $Ga_2O_3$ implanted by $Eu^+$ to different fluences. The profiles of defects and Eu atoms calculated using the SRIM code are also shown (in arbitrary units) as well as the Eu-profile extracted from the RBS spectra of the sample implanted with $D = 4\times10^{15}$ $cm^{-2}$. Reprinted with permission from [59] Copyright (2014) Society of Photo-Optical Instrumentation Engineers (SPIE).

These regularities are similar to those observed in the case of irradiation of GaN with heavy ions[79 80]: a two-peak distribution of damage and an increase in the thickness of the near-surface amorphized layer with increasing dose were also observed. The authors of [79 80] attribute the presence of the near-surface peak to the diffusion of mobile point defects from the bulk peak to the surface, or to the boundary of the amorphous region (after amorphization near the surface). This explanation later supported by [60] may apparently be valid also for $\beta$-$Ga_2O_3$. As noted in [79 80], the presence of two damage maxima in the case of GaN is typical of heavy ions (the $Eu^+$ used in Ref. [59] also corresponds to the type of heavy ions). Possibly, in the case of GaN and $\beta$-$Ga_2O_3$ (-201), the drift of point defects in the field of elastic stresses plays a role: the latter grows with increasing ion mass.

Interestingly, as seen in Figure 7, the tendency to saturation of damage is noticeable at certain depth. Another interesting feature observed in Ref. [59] is the non-monotonic dose dependence of RBS/C yield for high depths (deeper the implanted layer). The RBS/C level at these depths is significantly lower at doses at which amorphization occurs. The authors of Ref. [59] explain this by the fact that, while extended defects appear at comparatively low doses, causing dechanneling of $He^+$ ions, such defects are absent upon amorphization.

The study of annealing effect on the distribution of damage in β-$Ga_2O_3$ implanted by $Eu^+$ ions ($4\times10^{15}$ $cm^{-2}$)[59] also revealed a number of interesting features. RTA was carried out in an argon flow within the temperature range $T_{ann}$ = 500 – 1200 °C. After annealing at 500 °C, there is a slight decrease of damage at all depths (except for the tail section, where the occurrence of extended defects is assumed). A more significant recovery of the crystalline quality occurs at $T_{ann}$ = 700 °C and $T_{ann}$ = 900 °C when the distribution of damage has a two-peak character. Apparently at such temperatures a fraction of defect complexes decomposes into simpler mobile defects, which diffuse to the surface, while other fraction interacts with the Eu atoms and forms impurity-defect complexes. It can be noted that at $T_{ann}$ = 700 – 1000 °C the damage increases with temperature at the depths where the maximum concentration of Eu is located. This is consistent with the authors' assumption that the defects and Eu impurity stabilize each other. Finally, at $T_{ann}$ = 1100 °C, an almost complete recovery of the crystalline quality of β-$Ga_2O_3$ takes place and this is accompanied by a redistribution of Eu atoms – they move toward the surface.

In Ref. [60], the RBS/C method was used to study the effect of temperature during the implantation ($T_{imp}$) on the crystalline quality of β-$Ga_2O_3$ (-201), as well as the impact of post-implantation annealing after “hot” implantation by $Eu^+$ ($E$ = 300 keV, $D = 1\times10^{15}$ см$^{-2}$). It is demonstrated that the crystalline quality non-monotonically depends on $T_{imp}$ and $T_{ann}$. At $T_{imp}$ = 400 ° and $T_{imp}$ = 600 °C, the quality is significantly higher than that at $T_{imp}$ = 20 °C, while it decreases with elevation of $T_{imp}$ to 800 °C, and increases again only at $T_{imp}$ = 1000 °C. A similar dependence was observed for the annealing of samples irradiated at $T_{imp}$ = 300 ° and $T_{imp}$ = 600 °C in the range of $T_{ann}$ = 700 – 1000 °C. In this case, as the $T_{ann}$ increases, the fraction of Eu atoms replacing lattice sites decreases. This behavior additionally confirms the conclusions of Ref. [59], according to which defects and impurities mutually stabilize each other. Apparently, the interaction between them may be enhanced by local elastic stresses around impurity atoms. As in Ref. [73], the RBS/C data presented in Ref. [60] indicate the existence of extended defects.

The TEM and SEM methods were applied in Ref. [82] to investigate thermally evaporated $Ga_2O_3$ nanowires irradiated with $Eu^{3+}$ and $Gd^{3+}$ ions. It is found by SEM that nanowires have a complex shape and are oriented randomly in projection onto the substrate plane. The presence of an amorphous phase after irradiation is proven by TEM. A correlation between the structural changes (revealed by TEM) and CL is established. The recovery of crystalline quality upon annealing was investigated by Raman spectroscopy which demonstrated the recovery started at $T_{ann}$ = 500 °C and finished at 1100 °C. Study on β-$Ga_2O_3$ nanowires irradiated with $Eu^+$, $Gd^+$, and ions of some transition metals was performed in Refs. [62 63 83], where similar results were reported.

In Ref. [84], a detailed electron microscopy examination of epitaxial β-$Ga_2O_3$ layer irradiated by $Ge^+$ ions with doses and energies of $3\times10^{13}$ $cm^{-2}$/60 keV, $5\times10^{13}$ $cm^{-2}$/100 keV and $7\times10^{13}$ $cm^{-2}$/200 keV was performed. It is found out the damaged layer undergoes a phase transition from the β-phase to the k-phase during irradiation. Post-implantation annealing at 1150 °C for 60 s in $O_2$ atmosphere leads to the reverse transformation of k-phase into β-phase, except for a thin layer (~ 17 nm), which keeps the structure of the k-phase. In addition, a change of oxygen environment extended deeper than the implanted zone was discovered. These results show that the conclusions about damage accumulation and amorphization, which are based on RBS-C data, should be proven by direct structural methods, such as electron microscopy or XRD. In addition, they show the need for a thorough study of the starting material and the irradiation conditions influences, as well as the conditions of post-implantation annealing, on the nature of damage and structural-phase transformations under ion irradiation using various methods of analysis.

Let us now consider the investigations of the second type, where the point defects generated by ion irradiation are considered. In Ref. [85], the samples of β-$Ga_2O_3$ were irradiated by $O^+$ ions with $E$ = 25 MeV and $D$ = $1.5\times10^{11}$, $5\times10^{11}$ and $1.5\times10^{12}$ $cm^{-2}$. The main information was obtained by PL spectroscopy. It is established that the PL spectra of irradiated samples are similar to the spectra of non-irradiated ones, however the emission intensity is significantly higher. With increasing dose, the PL intensity decreases, apparently due to the accumulation of defects serving as the centers of nonradiative recombination. Considering the PL spectroscopy data, as well as the results of *ab initio* calculations[86], the authors identified the deconvoluted PL peaks with divacancies ($V_{Ga}$ + $V_O$), gallium vacancies ($V_{Ga}$), and interstitial oxygen atoms ($O_i$). The PL in this work was excited at a photon energy of 3.8 eV, which is lower than the band gap of β-$Ga_2O_3$. The peak at 3.8 eV was found in the photoluminescence excitation (PLE) spectrum.

In Ref. [87], β-$Ga_2O_3$ layers with an initial electron concentration of $4.8\times10^{17}$ $cm^{-3}$ grown on MgO substrate were irradiated at 35 K by $C^+$ ions with $E$ = 1.5 MeV and various doses up to $10^{14}$ $cm^{-2}$. The resistivity of films increased by more than 8 orders of magnitude, which was explained by the generation of defects that introduced deep levels into the band gap. The quantitative interpretation of the results was carried out on the basis of charge neutrality level (CNL) theory[88] (see also[89] [90] [91]).

The formation of defects under high-energy irradiation with light ions was studied in Refs. [92] [93]. Epitaxial β-$Ga_2O_3$ (010) layers with a Si concentration of ~ $10^{16}$ $cm^{-2}$ were irradiated with protons with $E$ = 10 and $E$ = 20 MeV at $D \approx 10^{14}$ $cm^{-2}$. The samples were characterized by the C-V method, C-V profiling under illumination, photocapacitance (PC), DLTS with electrical and optical excitations, as well as by deep level optical spectroscopy (DLOS). It is found that irradiation introduces into the bandgap several deep levels (traps) located both in the upper half of the band gap ($E_c$ – 0.6 eV, $E_c$ – 0.75 eV, $E_c$ – 1.05 eV, $E_c$ – 2.16 eV), and in its lower half ($E_v$ + 0.2 eV, $E_v$ + 0.4 eV, $E_v$ + 1.3 eV). The level ($E_v$ +1.3 eV) is attributed to gallium vacancies, while the levels ($E_c$ – 1.2 eV) and ($E_c$ – 2.3 eV) are close to the charge transfer levels of two types of oxygen vacancies predicted by the first-principle calculation in Ref. [17]. The authors associate the level ($E_v$ + 0.2 eV) with the STH (self-trapped hole) state. (After being trapped at the STH level the hole should become immobile.). According to DFT calculations [16], the transition of a hole from the valence band to the STH level and thus their immobilization is energetically favorable process and this is one of the main factors making it difficult to obtain *p*-type β-$Ga_2O_3$[3]. However, the authors of Refs. [92] [93] and Ref. [42] come to the conclusion that holes generated by light or by electron beam may be mobile. The conclusion about the existence of β-$Ga_2O_3$ with mobile holes, i.e. having *p*-type conductivity, was also made in Refs.[94] [95] [42]. Note, however, that the conclusion concerning the excited holes cannot be automatically transferred to the case of an equilibrium state of β-$Ga_2O_3$ doped with acceptor impurities.

In Ref. [19] single crystals and epitaxial layers of β-$Ga_2O_3$ were irradiated with protons at $E$ = 0.6 and $E$ = 1.9 MeV and $D = 5 \times 10^9 - 6 \times 10^{13}$ $cm^{-2}$. The irradiated samples with Schottky barrier were studied by DLTS and C-V methods, as well as by measuring the capacity recovery during heating. To interpret the obtained results, a series of *ab initio* calculations of the energy levels of various types of elementary point defects and complexes were performed. Some of the complexes contain hydrogen. In addition, the barriers for defect migration and recombination (trapping) were calculated. It was argued that the identification of experimental data of electron transitions requires taking into account all these factors, as well as shift of Fermi level and its pinning. Therefore, not only statistics, but also defect kinetics and temperature factors are important. The conclusion was made that the compensation of electronic conductivity under proton irradiation could be explained by the presence of Ga interstitials, gallium vacancies, and $Ga_O$ antisites, and that the migration and subsequent passivation of $V_{Ga}$ with hydrogen might be responsible for the thermal recovery process.

Electrical and optical methods provide information on the energy levels, but they do not provide information on the nature of defects. More informative for paramagnetic point defects is the method of EPR used in Ref. [34]. The UID β-$Ga_2O_3$ samples with a carrier concentration of $2\times10^{17}$ $cm^{-3}$ and semi-insulating β-$Ga_2O_3$ samples doped with Fe were irradiated by protons at $E$ = 12 MeV and $D = 10^{16}$ $cm^{-2}$. It is concluded that the assumption on the presence of single gallium vacancies in the irradiated samples is inconsistent with experimental data and that more probable is the $V_{Ga}$-$Ga_i$-$V_{Ga}$ complex predicted in Ref. [15].

A special type of damage is generated during irradiation by *swift* heavy ions, when the formation of ion tracks occurs. The tracks can be considered as cylindrical straight regions around the ion path when the energy $S_e$ deposited by the ion exceeds the value required for melting of the material. The rapid quenching freezes the defects resulting in formation of cylindrical regions with significantly changed atomic density and structure[96 97 98]. For some materials, the ion tracks may be amorphous[99], while for others the swift ion irradiation induces the recrystallization of an amorphous matrix[100] or forms the crystalline tracks with different density and structure than that of surrounding matrix.

The irradiation of β-$Ga_2O_3$ by swift ions was investigated also in Ref. [101] and Ref. [102]. In Ref. [101] the case of irradiation by $Au^+$ at $E$ = 946 MeV and $D$ up to $1\times10^{13}$ $cm^{-2}$ was considered. Using the XRD method, the amorphous structure of tracks was established. Their diameter was estimated to be 8.3 nm using the model of "inelastic thermal spike" (i-TS)[103 104 105]. In Ref. [102], the TEM method was utilized to study the tracks formed in β-$Ga_2O_3$ (100) under irradiation by swift ions of $^{86}$Kr и $^{181}$Ta. The authors confirmed the amorphous structure of the tracks and found that with an increase in electron losses $S_e$ from 18.3 to 41.8 keV / nm, the average track diameter increased from 2.2 to 8.8 nm. The data on the track diameters and the $S_e$ threshold value for the track formation were in good agreement with the i-TS model.

# V. $Ga_2O_3$-BASED DEVICES

## *A. Power electronic devices: MOSFET and SBD*

The need to improve the efficiency of communication systems and some types of electrical equipment, such as inductive motor controllers and power supplies, requires the development of power electronics components with the parameters unattainable for previously used semiconductors. As a base material for such devices, β-$Ga_2O_3$ is of particular interest [3 7] due to its large band gap which provides a high breakdown voltage $V_{br}$, the ability to withstand high temperatures, as well as the possibility to change the conductivity over a very wide range by doping, etc. When choosing a material for the power devices fabrication, a very important criterion is the Baliga figure of merit (BFOM)[106], which determines the power losses during device operation: the higher the BFOM is, the lower power losses are. The relation BFOM = $V_{br}^2/R_{on}$ is based on the assumption that power losses are due to energy dissipation in the "on-state" resistance $R_{on}$. (It should be noted that BFOM can be used for devices operating at low frequencies or at direct current (DC)[106]).

Ion implantation of β-$Ga_2O_3$ was investigated for the development of two types of power devices – MOSFET and SBD. In particular, the implantation of $Si^+$ ions into β-$Ga_2O_3$ layers was used to form ohmic contacts for SBD[50 107 46], contact regions of the source/drain and the channels of MOSFET[108 109 110 111 112 113 114 115 116 117]. By $Mg^+$ or $N^+$ implantation into $\beta$-$Ga_2O_3$ layers, a Current Blocking Layer (CBL) in vertical type MOSFET devices was created[114 115 113], as well as high-resistance regions and guard rings near the SBD anode[107 118]. In Ref. [119], the possibility to create both low-resistance regions for the formation of ohmic contacts and high-resistance layers near the edges of the SBD anode by $Ar^+$ implantation was shown. For the formation of MOSFETs on heterogeneous structures, a method of exfoliation and bonding was developed, based on irradiation with $H^+$ and $Ar^+$ [120 121]. Let's consider these works in more detail.

From the analysis of data reported [50 107 46 108 110 122 112 113 114 115], it is possible to trace the evolution of the purposes for which the ion implantation was used over the past few years. In the early work of this group [46], the electrical properties of $\beta$-$Ga_2O_3$ layers upon implantation of $Si^+$ ions, followed by annealing were investigated, and the possibility of creating ohmic contacts by $Si^+$ implantation into the near-contact regions of devices was for the first time shown (Figure 8a). Multiple $Si^+$ ion implantation was performed at energies $E$ = 10 – 175 keV. The total implantation dose $D$ varied from $2\times10^{14}$ cm$^{-2}$ to $2\times10^{15}$ cm$^{-2}$. The activation annealing was carried out at temperatures of 900 – 1000 °C in an $N_2$ atmosphere for 30 minutes. Ti (50 nm)/ Au (300 nm) layers were used as ohmic contacts. At the concentration of implanted Si of $5\times10^{19}$ cm$^{-3}$, the authors achieved the specific contact resistance $\rho_c = 4.6\times10^{-6}$ $\Omega\times$cm$^2$, which was an order of magnitude lower than in the case of Sn diffusion[123] and lower than in the works[124 125], where $Si^+$ implantation was used at $E$ = 30 keV and $D = 1\times10^{15}$ cm$^{-2}$, activation annealing was carried out at a temperature of 950 °C, and ohmic contacts were created by the deposition of ITO (Indium Tin Oxide) or AZO (Aluminum Zinc Oxide) films with subsequent deposition of a Ti/Au film. (In Refs. [124 125], the values of $\rho_c$ were $6.3\times10^{-5}$ $\Omega\times$cm$^2$ and $2.8\times10^{-5}$ $\Omega\times$cm$^2$, respectively).

In subsequent works [108 109 110 122 112], ion implantation was used in the development of a depleted-mode MOSFET (D-mode MOSFET). At the initial stage, $Si^+$ implantation was used only for the formation of ohmic $n^{++}$ contact regions to the source/drain [108], while the channel was created by Sn doping during the growth of $Ga_2O_3$ film. $Si^+$ ions were implanted to a depth of 150 nm. The Si concentration in the implanted regions was $5\times10^{19}$ cm$^{-3}$. The activation annealing was carried out at a temperature of 925 °C in an $N_2$ atmosphere for 30 minutes. The $\rho_c$ value was $8.1\times10^{-6}$ $\Omega\times$cm$^2$. Since the Si concentration near the surface was lower than in the region of the distribution maximum (due to the Gaussian profile of the impurity during ion implantation), a 13-nm-deep recess in the source and drain regions was made before metallization by reactive ion etching. The Ti(20 nm)/Au(230 nm) films were used as contacts. In this way, devices were created with the following characteristics: a gate length $L_g$ = 2 μm; drain current $I_d$ = 39 mA/mm at gate voltage $V_g$ = +4 V; breakdown voltage $V_{br}$ = 404 V at $V_g$ = –20 V; the ratio of currents in open and closed states $I_{don}/I_{doff} > 10^{10}$. The MOSFET characteristics remain stable over the operating temperature range (20 – 250 °C).

In continuation of this series of works, $Si^+$ implantation was used not only to improve the ohmic contacts but also to form the MOSFET channels. For this purpose, a 300-nm-deep *n*-type region was created by implantation of $Si^+$ into a β-$Ga_2O_3$ film grown by molecular beam epitaxy on a semi-insulating gallium oxide substrate (exhibited in Figure 8b)[109]. Implantation conditions during the channel formation[109 110] were: $E = 10 – 330$ keV; $D = 1.1\times10^{13}$ $cm^{-2}$. Annealing was carried out under the conditions specified in Ref. [108]. The average concentration of charge carriers in the channel was $3\times10^{17}$ $cm^{-3}$. The authors also noticed the advantage of the channels formation by $Si^+$ implantation as compared to Sn doping during the epitaxy of $Ga_2O_3$ layer when the Sn impurity segregation took place[112]. The creation of a channel by $Si^+$ ion implantation made it possible to improve the characteristics of the MOSFET in comparison with the work [108] and to obtain a MOSFET with the following parameters: $I_d$ = 65 mA/mm at $V_g$ = +6 V; $V_{br}$ = 415 V at $V_g$ = – 30 V; $I_{don}/I_{doff} > 10^{10}$.

In Refs. [108 109], a β-$Ga_2O_3$ wafer with a Fe concentration of ~ $10^{18}$ $cm^{-3}$ was used as a semi-insulating substrate. However, subsequent studies[110] showed that in this case there was an enhanced diffusion of Fe atoms from the semi-insulating substrate into the implanted semiconductor layer due to defects formed upon $Si^+$ implantation. This diffusion leaded to deterioration in the MOSFET performance. However, the formation of a buffer UID β-$Ga_2O_3$ layer between a semi-insulating substrate and a $Si^+$ implanted channel (Figure 8c) effectively eliminated the penetration of Fe into the channel region[110].

In Ref. [111], an additional Field-Plate (FP) electrode (FP-MOSFET mode, Figure 8d) was introduced into the device design, which made it possible to achieve the following device parameters: $V_{br}$ = 755 V at $V_g$ = – 55 V; $I_d$ = 78 mA/mm at $V_g$ = +4 V. It should be noted[111] that there was practically no variation of the drain current with the stability of characteristics up to operating temperatures of 300 °C.

So far D-mode MOSFETs were discussed. However, the Enhanced-mode (E-mode) MOSFETs are widely used in digital and power electronics. In Ref. [113], $Si^+$ implantation was used for the development of devices with $n^{++}$ contact regions of the source/drain (Figure 9). The devices were characterized by a low parasitic resistance between the source and drain (2.2 Ω×mm) and a small specific contact resistivity ($\rho_c = 7.5\times10^{-6}$ $\Omega\times cm^2$).

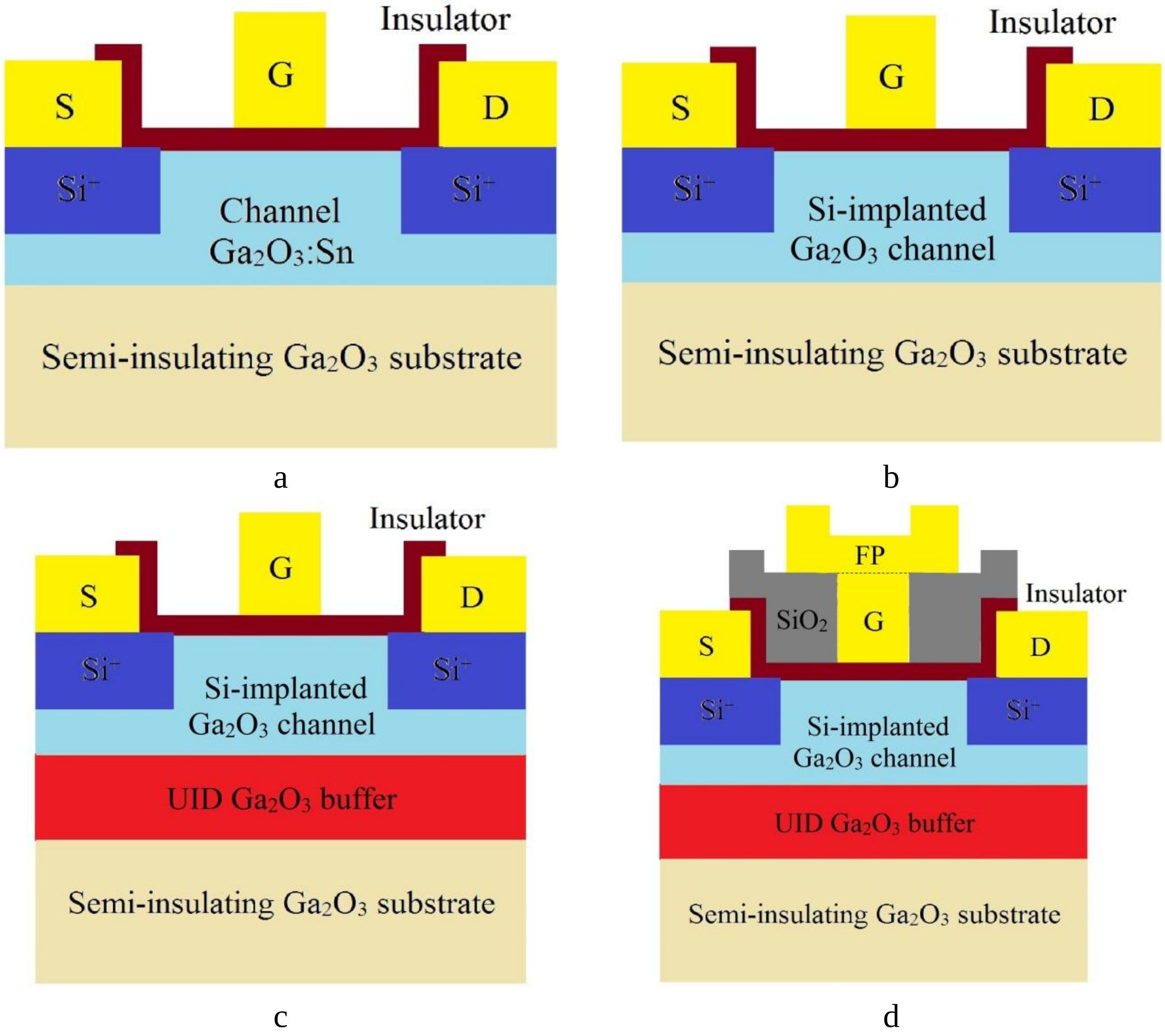

Figure 8. D-mode MOSFET designs with regions created by $Si^{+}$ implantation: (a) with $n^{++}$ near-contact source/drain regions; (b) the same as (a), but with a channel formed by $Si^{+}$ ion implantation; (c) the same as (b), but with an additional buffer layer; (d) the same as (c), but with an additional FP electrode.

Even better characteristics of MOSFETs with $n^{++}$ source and drain contact regions formed by $Si^{+}$ implantation were achieved in Refs. [116 117 126], where a higher implantation dose ($1.5\times10^{15}$ $cm^{-2}$) was used. Ion energies during multiple implantations were 10, 30, 60, and 100 keV. The total implantation depth was 210 nm, and the average silicon concentration was – $10^{20}$ $cm^{-3}$. For the D-mode FP-MOSFET, the following parameters were achieved: $V_{br}$ = 720 V, when the devices operated in air, and $V_{br}$ = 2360 V, when operated in a mixture of inert gas with fluorine. In Ref. [126], the values of $V_{br}$ = 2900 V and BFOM = 182 $MW\times cm^{-2}$ were obtained for devices of this type (at the distance between the gate and drain $L_{GD}$ = 17.8 μm) using a T-shaped gate and $Al_2O_3/HfO_2$ layers as a gate dielectric. For the E-mode FP-MOSFET[127], the use of $Si^{+}$ implantation for the source and drain formation made it possible to achieve the following values of $V_{br}$, BFOM, and contact resistance $R_c$: 3000 V; 94 $MW\times cm^{-2}$; 2 Ω ×mm, respectively. Finally, in Ref. [128], $Si^+$ implantation was used to form a source and drain, a self-alighned gate, and a thin channel (22 nm). A record value of the maximum transconductance of 35 mS/mm was obtained. The $R_c$ value was 1.5 Ω ×mm.

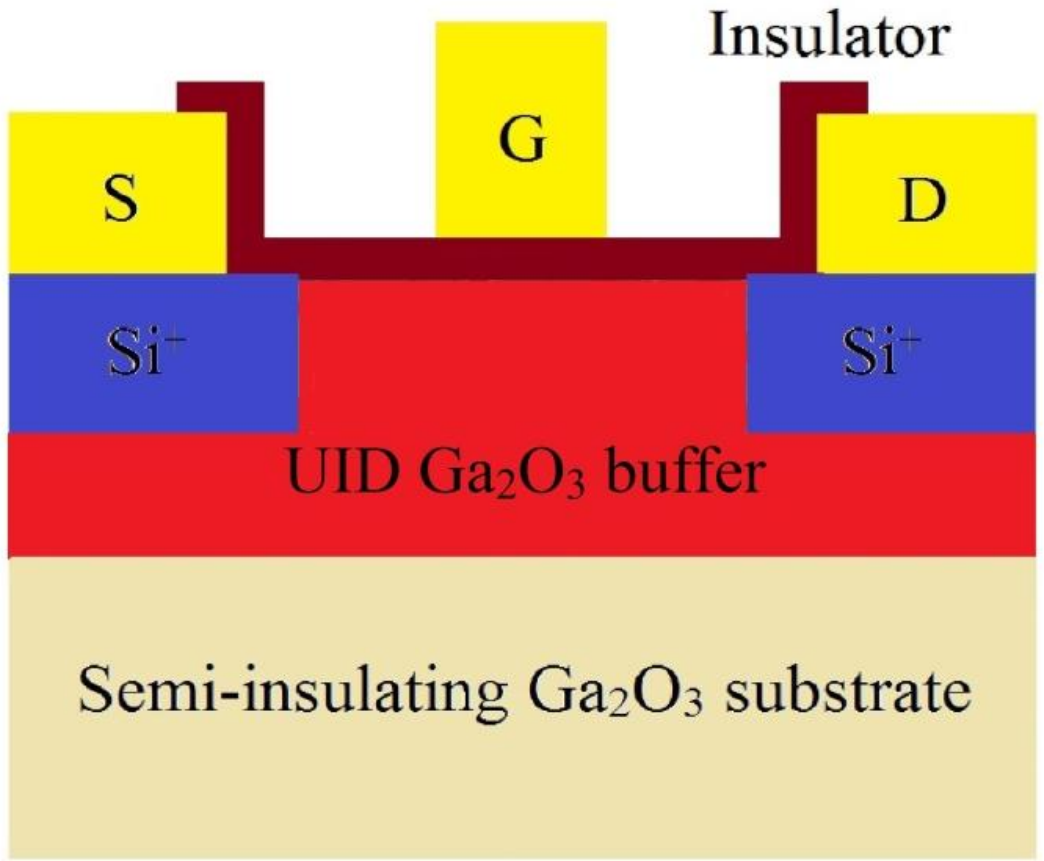

Figure 9. E-mode MOSFET design with $n^{++}$ ion-implanted source/drain contact regions.

The β-$Ga_2O_3$ disadvantage is its very low thermal conductivity, which creates certain problems in the development of power electronic devices based on this material. To improve the conditions for heat dissipation, it seems that it would be sufficient to deposit thin films of gallium oxide on materials with a higher thermal conductivity, e.g. on diamond or SiC [120 121]. However, due to the large lattice mismatch, the synthesis of β-$Ga_2O_3$ films with high-quality on such substrates is problematic. In works [119 120 121], the ion-cutting (exfoliation) technique was used to eliminate this disadvantage. Bulk β-$Ga_2O_3$ was implanted with high-dose $H^+$. After that, an interface layer was formed on the wafer surface by deposition of $Al_2O_3$ or amorphization of β-$Ga_2O_3$ by irradiation with $Ar^+$ ions, and the resulting structure was bonded with Si or SiC substrate. During the subsequent annealing, the implanted hydrogen migrated and accumulated at a certain depth, creating a high level of stress. As a result of this stress, the exfoliation of β-$Ga_2O_3$ thin layer occurred. Since the conductivity of $Ga_2O_3$ film was deteriorated upon $H^+$ implantation, the obtained heterogeneous structure was doped by $Si^+$ implantation at $E$ = 15 keV and $D$ = $5\times10^{13}$ $cm^{-2}$, after which D-mode and E-mode MOSFETs were fabricated on the gallium oxide film. The source and drain areas were doped using additional silicon implantation. The authors noted the high stability of the obtained devices characteristics in comparison with the devices fabricated on β-$Ga_2O_3$ substrates. The breakdown voltage $V_{br}$ was 570 V at 300 K and increased to 605 V at 500 K.

Now, let us consider the use of $Mg^+$ and $N^+$ acceptor impurities implantation for the development of power electronic devices. In Ref. [50], the effect of $Mg^+$ and $N^+$ implantation, as well as the subsequent activation annealing temperature was investigated in order to establish which of the conditions were more efficient for the development of high-voltage SBDs. The conditions for $Mg^+$ implantation into β-$Ga_2O_3$ (grown by the EFG method) were the following: $E$ = 560 keV; $D = 6\times10^{14}$ $cm^{-2}$. The magnesium concentration at a depth of 0.5 – 0.6 μm was $1.5 \times 10^{19}$ $cm^{-3}$. Subsequent annealing was carried out at temperatures of 600 – 1000 °C for 30 minutes in $N_2$ atmosphere. $N^+$ implantation was carried out into UID β-$Ga_2O_3$ at E = 480 keV and D = $4\times10^{13}$ $cm^{-2}$. The nitrogen concentration at a depth of 0.5 – 0.6 μm was $1.5\times10^{18}$ $cm^{-3}$, and post-implantation annealing was performed at 800 – 1200 °C. Due to the insufficiently high concentration of charge carriers in the UID β-$Ga_2O_3$ substrate, ohmic contacts were formed using $Si^+$ implantation. It is found that, in contrast to the layers with $N^+$ implantation, defect clusters are formed in β-$Ga_2O_3$ layers implanted by $Mg^+$. As a result, with an increase in annealing temperature, the leakage current increases. For the samples with $N^+$ implantation, the leakage current, on the contrary, decreases. Therefore, using $N^+$ ions gives better results.

In Refs. [114] [115] [113], the results of Ref. [50] are used to design a vertical mode MOSFET (Figure 10), in which the gate and source are located on one side of the crystal, and the drain is formed on the opposite side. It is noted [113] that, for high voltage and high power devices, vertical mode is highly desirable since it allows superior field termination and current drives. For such MOSFETs, it becomes necessary to form areas of charge carrier drift by creating a current blocking layer (CBL). This layer was formed by $Mg^+$ and $N^+$ ion implantation into gallium oxide layers grown by the HVPE method (Figure 10a). The conditions for $Mg^+$ implantation[114] were as follows: $E$ = 560 keV; $D$ = $8\times10^{12}$ $cm^{-2}$, the conditions for $N^+$ implantation were the same as in Ref. [50]. The magnesium and nitrogen concentrations at a depth of 0.5 – 0.6 μm were $2\times10^{17}$ $cm^{-3}$ and $1.5\times10^{18}$ $cm^{-3}$, respectively. After the activation annealing for 30 minutes in $N_2$ at 1000 °C for $Mg^+$ [114] and 1100 °C for $N^+$ [115], these impurities served as compensating acceptors, and significantly reduced the electron concentration in CBL. Transistor structures with CBL formed by $Mg^+$ implantation were characterized by a significant leakage current (exceeding 10 A/$cm^2$ at a voltage of 1 V). The reason for this was explained by the Mg diffusion confirmed by the SIMS. The nitrogen has a significantly lower diffusion coefficient, which makes it possible to use higher annealing temperatures required for the efficient activation of embedded ions[115]. Thus, $N^+$ implantation for the CBL formation is preferred. It should be noted that, according to Refs. [114] [115], the characteristics of vertical mode MOSFETs were still inferior compared to lateral mode.

For the development of an E-mode MOSFET of vertical type[113], in addition to the formation of $n^+$ channel by $Si^+$ implantation and $n^{++}$ contact regions and CBL by $N^+$ implantation, an access region was formed that differed from the contact regions by a higher concentration of $Si^+$ (Figure 10b). The characteristics of the E-mode MOSFET of vertical type in Ref.[113] were also inferior to the characteristics of the MOSFET of lateral type, but the authors of Ref. [113] expect that, with improved dielectric quality and optimized doping schemes; their work promises a transformational impact.

However ,in Ref.[129] record values of $V_{br}$ = 2633 V and a BFOM = 280 MW×cm$^{-2}$ for the vertical multi-fin E-mode MOSFET among all types of power $Ga_2O_3$ MOSFETs, in which ion implantation was used to create ohmic contacts to source, were demonstrated. The authors noted the effect of post-deposition annealing at 350 °C for 1 min under $N_2$ to achieve a record low specific on-resistance = 5.2 mΩ×cm$^2$ and a high channel mobility of ~ 130 cm$^2$/(V×s).

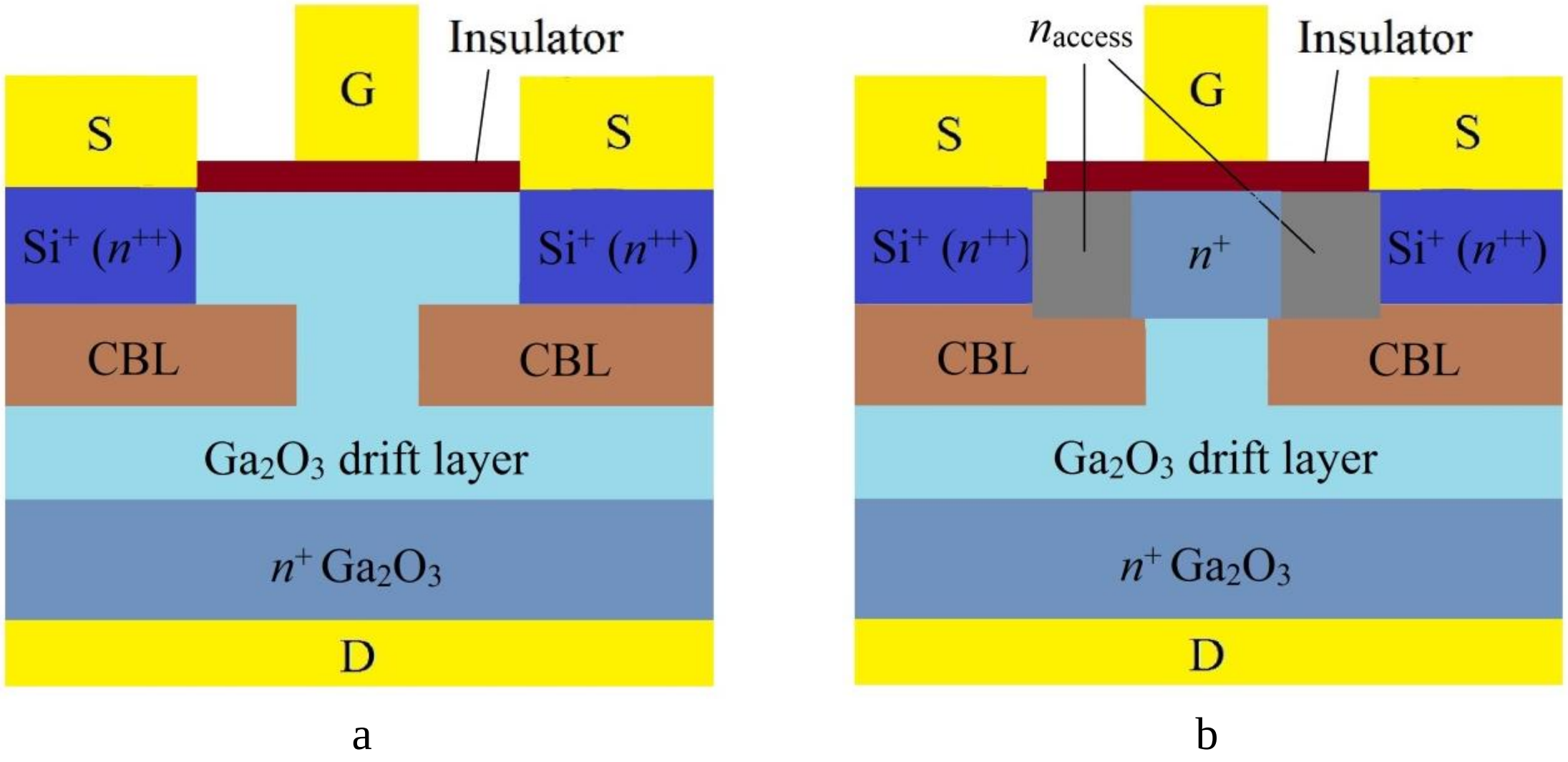

Figure 10. Vertical type MOSFET designs with active regions created by ion implantation: (a) with $n^{++}$ contact source/drain regions and CBL region doped by $N^+$ or $Mg^+$; (b) the same as (a), but with access area.

Let us now consider the works, in which ion implantation was used to create β-$Ga_2O_3$-based power SBDs. In Refs. [107, 118], it was found that the use of implantation for the formation of high-resistance regions or guard rings (G.R.) at the anode allows reducing the reverse current and significantly increasing $V_{br}$ in the vertical mode SBD (Figure 11). In Ref. [107], the G.R. was formed by $N^+$ ion implantation to the 0.8 μm depth with a concentration of 1.0×10$^{17}$ cm$^{-3}$. The activation annealing was carried out at 1100 °C for 30 minutes in an $N_2$ atmosphere. The G.R. formation increased $V_{br}$ from 750 to 860 V. The authors used G.R. also to create an SBD with a field-plated electrode and found that in this case $V_{br}$ increased from 1380 to 1430 V. In Ref. [118], $Mg^+$ implantation with the following energy/dose ratios: 50 keV/1.4×10$^{14}$ cm$^{-2}$; 125 keV/2×10$^{14}$ cm$^{-2}$ and 250 keV/9.8×10$^{14}$ cm$^{-2}$ was used to create a high-resistance region in SBD. The doping region depth was 0.8 μm, and the implanted impurity concentration was (2 – 3)×10$^{19}$ cm$^{-3}$. Annealing after implantation was not carried out, so the increase in resistivity was associated apparently not with the Mg acceptor properties, but with the donor compensation by radiation defects. The use of high-resistance region reduced the reverse current and increased $V_{br}$ from 500 to 1550 V.

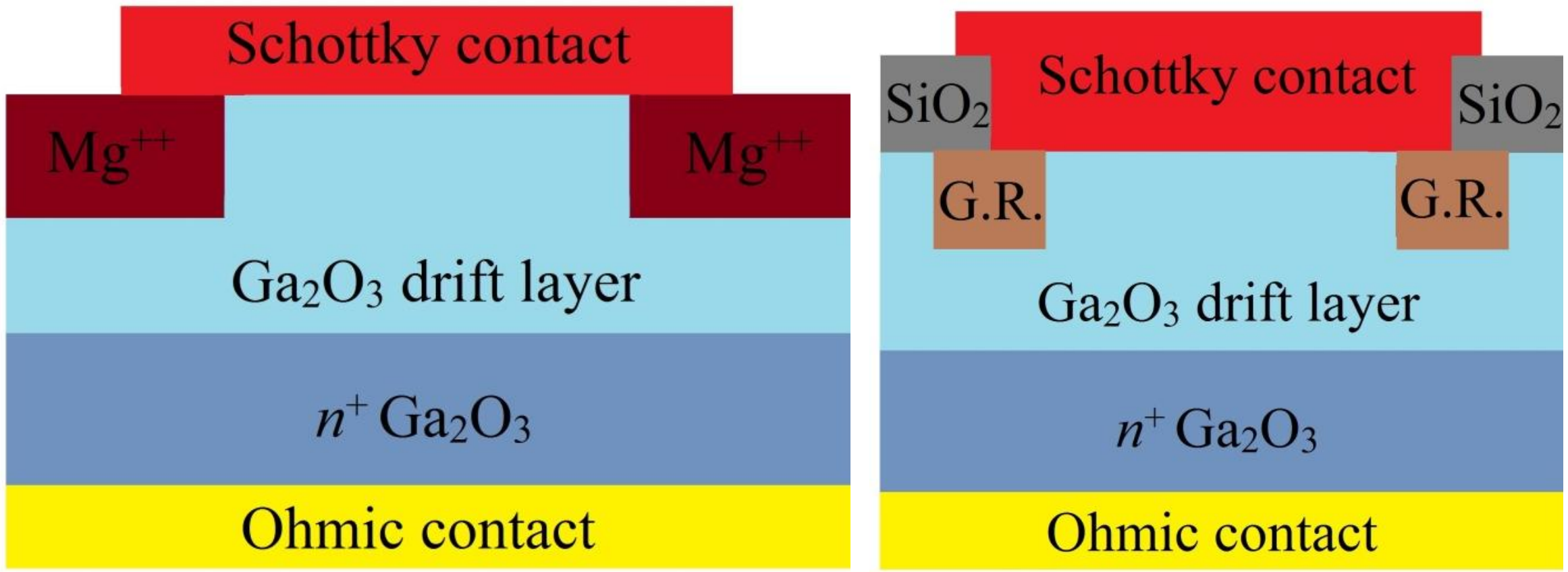

a b

Figure 11. SBD structures based on β-$Ga_2O_3$ with regions formed by $N^+$ or $Mg^+$ implantation: (a) with a high-resistance region near the anode; (b) with G.R.

In Ref. [119], as mentioned above, the possibility of using $Ar^+$ ion implantation to form both low-resistance regions under ohmic contacts and a high-resistance region at the edges of the anode was shown, as well. The implantation conditions used for the ohmic contacts formation were the following: $E$ = 50 keV; $D = 2.5\times10^{14}$ cm$^{-2}$. Annealing after irradiation was carried out at $T_{ann}$ = 950 °C for 60 minutes in a $N_2$ atmosphere. To form the high-resistance region, two irradiation doses of $5\times10^{14}$ cm$^{-2}$ and $1\times10^{16}$ cm$^{-2}$ were used, and annealing was performed at $T_{ann}$ = 400 °C for 60 s in a $N_2$ atmosphere, too. Increasing the dose led to an increase in $V_{br}$ from 252 to 451 V. These devices demonstrated thermal stability of characteristics in the range from room temperature to 150 °C. Theoretical calculations aimed to reveal the role of defects formed upon $Ar^+$ irradiation are presented in Ref. [130].

The characteristics of β-$Ga_2O_3$-based MOSFETs and SBDs processed by ion implantation are summarized in Table 3. The developed devices demonstrate high breakdown voltages, low leakage currents, low resistance of ohmic contacts and stability of parameters in a wide temperature range. It is remarkable that almost every new publication reports new achievements in terms of device characteristics improvement. Apparently, the use of ion implantation has not yet reached the limits of its capabilities in improving the characteristics of these devices.

Table 3. β-$Ga_2O_3$-based MOSFET and SBD characteristics.

| Ions | Purpose of implantation | Type of device | Substrate | $V_{br}$, V | BFOM, MW×cm$^{-2}$ | $I_{on}$ / $I_{off}$ | $\rho_c$, | $V_{br}$, V |
|---|---|---|---|---|---|---|---|---|
| $Si^+$ | | Lateral D-mode MOSFET | – | 404 | – | >$10^{10}$ | 8.1×$10^{-6}$ | Ref. [108] |
| $Si^+$ | Ohmic contacts | Lateral D-mode MOSFET | – | 415 | – | >$10^{10}$ | 8×$10^{-6}$ | Ref. [109] |
| $Si^+$ | Ohmic contacts, channel | Lateral D-mode MOSFET | UID $Ga_2O_3$, buffer layer | 480 for air annealing, 1340 for fluorinert annealing | – | $10^5$ | – | Ref. [116] |
| $Si^+$ | Ohmic contacts | Lateral D-mode FP-MOSFET | UID $Ga_2O_3$, buffer layer | 755 | – | >$10^9$ | – | Ref. [111] |
| $Si^+$ | Ohmic contacts, channel | Lateral D-mode FP-MOSFET | UID $Ga_2O_3$, buffer layer | 720 for air annealing, 2360 for fluorine annealing | 8.8 | $10^5$ | – | Ref. [116] |
| $Si^+$ | Ohmic contacts | Lateral D-mode FP-MOSFET | UID $Ga_2O_3$, buffer layer | 480 | 50.4 | $10^5$ | 4.6×$10^{-3}$ | Ref. [117] |
| $Si^+$ | Ohmic contacts | Lateral D-mode FP-MOSFET | UID $Ga_2O_3$, buffer layer | 2900 for $L_{GD}$=17.8 μm, 1400 for $L_{GD}$=4.8 μm | 182 for $L_{GD}$=17.8 μm 277 for $L_{GD}$=4.8 μm | $10^9$ | – | Ref. [126] |
| $Si^+$ | Ohmic contacts | Lateral E-mode FP-MOSFET | UID $Ga_2O_3$, buffer layer | 3000 | 94 | $10^8$ | – | Ref. [127] |
| $Si^+$ | Ohmic contacts | Lateral E-mode MOSFET | UID $Ga_2O_3$, buffer layer | – | – | $9\times10^5$ | $7.5\times10^{-6}$ | Ref. [113] |
| $Si^+$ | Ohmic contacts | Lateral D-mode MOSFET | – | – | – | $10^8$ | – | Ref. [128] |
| $Si^+$ | Ohmic contacts | Lateral D-mode and E-mode MOSFET | Heterogeneous structure | 570, 605 | – | $10^7$ | – | Refs.[120,121] |
| $H^+$ | Ohmic contacts to increase the conductance of $Ga_2O_3$ | | | | | | | |
| $Ar^+$ | Exfoliation of β-Ga2O3 thin film | | | | | | | |
| $Si^+$ | Bonding interface layer | Vertical D-mode FP-MOSFET | – | – | – | $10^8$ | – | Ref. [114] |
| $N^+$ | Ohmic contacts, channel | | | | | | | |
| $Si^+$ | CBL | Vertical D-mode FP-MOSFET | – | – | – | – | – | Ref. [115] |
| $Mg^+$ | Ohmic contacts, channel | | | | | | | |
| $Si^+$ | CBL | Vertical E-mode MOSFET | – | 263 | 0.5 | $2\times10^7$ | – | Ref. [113] |
| $N^+$ | Ohmic contacts, channel, access region | | | | | | | |
| $Si^+$ | Ohmic contacts | Vertical multi-fin E-mode MOSFET | – | 2655 | 280 | $10^9$ | – | Ref.[129] |
| $N^+$ | CBL | SBD | – | 860 for G.R., 1430 for G.R. and FP electrode | 150 for G.R., 400 for G.R. and FP electrode | – | – | Ref. [107] |
| $Mg^+$ | G.R. | SBD | – | 1550 | 470 | – | – | Ref. [118] |
| $Ar^+$ | Edge termination | SBD | – | 451 | 61.5 | – | – | Ref. [119] |

## B. Solar-blind UV detectors

Photodetectors with selective sensitivity to ultraviolet (UV) radiation with $\lambda \leq 280$ nm are commonly called "solar-blind", as they can function in sunlight conditions[131]. They find applications in various fields, such as detection of laser UV radiation, detection of cruise missile tracks, investigations of the Earth's atmosphere (e.g., detection of ozone holes), astrophysics, space investigations, biology, medicine, etc.

Solar-blind UV detectors should have high values of photosensitivity $R_{ph}$, detectivity $D_{ph}$, high-performance, and low dark current $I_d$. The $R_{ph}$ value is determined from the expression:

$$R_{ph} = (I - I_d)/P, \qquad (3)$$

where $I$ is the total current when detector is exposed to radiation, and $P$ is the radiation power. The detectivity $D_{ph}$ characterizes the ability of device to detect small UV signals and is determined by the ratio[132]:

$$D_{ph} = [S/(2qI_d)]^{1/2} \times R_{ph}, \qquad (4)$$

where $S$ is the effective area of the photodetector, $q$ is the electron charge. The performance of detector is characterized by the time of photoresponse under pulsed radiation.

Recently, a fairly large number of articles have appeared devoted to the research and development of solar-blind photodetectors based on $\beta$-$Ga_2O_3$[131 132]. Most attention is paid to the detectors of metal/$\beta$-$Ga_2O_3$/metal (Me/$\beta$-$Ga_2O_3$/Me) type, where the metal is used to create an ohmic contact or a Schottky barrier. UV detectors with ohmic contacts are characterized by high sensitivity, but at the same time they have high values of the dark current[131]. The Schottky barrier detectors have lower dark currents and better response times due to the presence of the barrier electric field; however, they tend to have lower sensitivity than devices with ohmic contacts. It should be noted that high sensitivity is not always the main characteristic; for some applications, high performance is more important.

So far, only a few works have been devoted to the use of ion implantation for the creation and modification of the $\beta$-$Ga_2O_3$-based UV detectors properties. In Ref. [133], a $\beta$-$Ga_2O_3$ layer epitaxially grown by the MOCVD method on a sapphire (0001) substrate was implanted by $Si^+$ with an energy of 30 keV and a dose of $1\times10^{15}$ $cm^{-2}$, after which the samples were annealed at 900 °C for 30 – 120 s in Ar atmosphere. The depth corresponding to the maximum concentration of implanted ions was 25 nm. $Si^+$ implantation and an increase in the annealing duration from 30 to 90 s led to a significant decrease in the surface resistance of the film. Ti / Au films were used as ohmic contacts. The $R_{ph}$ value of the structures upon exposure to UV radiation with λ = 254 nm was 1.45 A/W. The samples response was characterized by significant inertia due to the presence of a deep level in the β-$Ga_2O_3$ band gap. Unfortunately, the authors did not compare the characteristics of the obtained detectors with the characteristics of detectors fabricated without $Si^+$ implantation. (It is worth noting that a decrease in the contact resistance due to implantation can lead to an increase in the dark current and, accordingly, a decrease in $D_{ph}$).

The report [133] was continued by the study of characteristics in a wide temperatures ($T$) range – from room temperature to 350 °C [134]. The near-surface region of β-$Ga_2O_3$ layer was implanted by $Si^+$ and then the samples were annealed under the conditions described above. The detectors showed stable performance over the entire studied temperature range, and at $T$ = 400 °C, the Ti/Au contacts degradation occurred. An increase in the *rejection ratio* ($I_{ph}$(λ = 254 nm) / $I_{ph}$(λ = 365 nm)) from 9 at room temperature to 216 at 350 °C was noted. (The *rejection ratio* parameter characterizes the degree of the detector suitability for the use as a "solar-blind" detector). With an increase in $T$ from room temperature to 350 °C, the difference between the photocurrent at λ = 254 nm and the dark current increased by about a factor of 7, and the $R_{ph}$ value increased from 5 to 36 A/W.

In Ref. [135], the sensitivity of $Ga_2O_3$ nanowires synthesized employing plasma immersion ion implantation (PIII) of GaAs with $C_2H_2^+$ ions, to pulsed nitrogen laser emission with a wavelength of 337 nm was studied. The lack of nanowires sensitivity to emission with λ = 488 nm was noted. It is concluded that the method is promising for UV detectors and for the fabrication of integrated optical circuits.

# VI. SUMMARY AND CONCLUSIONS

More than half a century of research and development in the field of semiconductor electronics shows that technologies relevant to industrial application hardly could be created without ion implantation due to its many advantages. This fully applies to the case of $Ga_2O_3$ as a material for advanced electronic and optoelectronic devices. Hence, it follows that research on ion implantation for this semiconductor should be given a scale comparable to those for the well-developed “classical” semiconductors.

Comprehensive considerations of all the aspects in this field is beyond the scope of the present review. Here, the topics that seem to be the most important and interesting are considered.

To date, the gallium oxide based MOSFET devices for power electronics have been demonstrated. Such devices can also serve as a basis for the development of *integrated* circuits operating at high temperatures. In order to meet the requirements for high performance and high degree of integration, a microminiaturization of circuits is

necessary – achieving very small sizes of the active and passive regions. From this point of view, ion implantation is the most appropriate technique. However, as the experience in the development of silicon integrated circuits shows, to implement this task, numerous studies are required. Among them are such problems as super-shallow doping (by implantation of low-energy ions), studying the regularities of radiation-stimulated diffusion and finding innovative ways to suppress it (for example, by implanting two and more impurities, irradiation with molecular and cluster ions, use of impurity gettering). Additional research is also required in the field of "defect engineering" – the regularities of accumulation and annealing of radiation defects, the interaction of impurities and defects, the effect of elastic stresses on diffusion, the influence of crystal surface and its orientation on the distribution of implanted atoms, etc. These problems are either not yet investigated for $Ga_2O_3$, or their study is in infancy state.

One of the significant drawbacks of $Ga_2O_3$, limiting its application, is the difficulty of producing *p*-type. This is due to the fundamental features of the energy structure of this material under the conditions close to thermodynamic equilibrium. A promising approach to search the way of solving this problem is to realize nonequilibrium, but relative stable (metastable) states with the help of ion implantation. Perhaps such an approach (using ion implantation), as the introduction of a "third player" into the game (the first two "players" are a semiconductor matrix and an impurity or defect) also would be fruitful. The third "player" can be, for example, a surface, the role of which increases in the case of using low-energy ions, or new ion-synthesized phases, including nanoinclusions, interfaces, etc. Ion synthesis of ternary compounds with the aim of modifying the energy structure of a semiconductor is also a promising area of research.

All these problems require not only extensive experimental research, but also the intensification of theoretical developments.

Although this review refers to the β-phase of $Ga_2O_3$, possible applications of gallium oxide are not limited to this phase. At present, there are prospects for synthesis of epitaxial layers of the α-phase (with a wider band gap than the β-phase). The problems with the α-phase ion implantation are likely to be more difficult due to its lower thermal stability. Just established formation of the k-phase by ion implantation [84] would be of large interest, as well.

Obviously, further researches in the area of $Ga_2O_3$ ion implantation is very important from both scientific and practical points of view.

# ACKNOWLEDGMENTS

The study was supported by the Lobachevsky University competitiveness program in the frame of 5-100 Russian Academic Excellence Project.

## DATA AVAILABILITY

The data that support the findings of this study are available from the corresponding author upon reasonable request.